\documentclass[]{spie}  

\usepackage{amsmath,amsfonts,amssymb}
\usepackage{graphicx}
\usepackage{multirow, makecell}
\usepackage{rotating}

\title{CAMELOT: Cubesats Applied for MEasuring and LOcalising Transients - Mission Overview}

\author[a,b,c]{Norbert Werner }
\author[a,d]{Jakub {\v R}{\'i}pa }
\author[e]{Andr\'as P\'al}
\author[c]{Masanori Ohno}
\author[f]{Norbert Tarcai}
\author[c]{Kento Torigoe}
\author[c]{Koji Tanaka}
\author[c]{Nagomi Uchida}
\author[e]{L\'aszl\'o M\'esz\'aros}
\author[g]{G\'abor Galg\'oczi}
\author[c]{Yasushi Fukazawa}
\author[c]{Tsunefumi Mizuno}
\author[c]{Hiromitsu Takahashi}
\author[h]{Kazuhiro Nakazawa}
\author[f]{Zsolt V{\'a}rhegyi}
\author[i]{Teruaki Enoto}
\author[j]{Hirokazu Odaka}
\author[k]{Yuto Ichinohe}
\author[g,l]{Zsolt Frei}
\author[e]{L\'aszl\'o Kiss}

\affil[a]{MTA-E\"ot\"vos University Lend\"ulet Hot Universe Research Group, P\'azm\'any P\'eter s\'et\'any 1/A, Budapest, 1117, Hungary}
\affil[b]{Department of Theoretical Physics and Astrophysics, Faculty of Science, Masaryk University, Kotl\'a\v{r}sk\'a 2, Brno, 611 37, Czech Republic }
\affil[c]{School of Science, Hiroshima University, 1-3-1 Kagamiyama, Higashi-Hiroshima, Japan }
\affil[d]{Charles University, Faculty of Mathematics and Physics, Astronomical Institute, V Hole\v{s}ovi\v{c}k\'ach 2, 180 00 Prague 8, Czech Republic }
\affil[e]{Konkoly Observatory of the Hungarian Academy of Sciences, Konkoly-Thege ut 15-17, Budapest, 1121, Hungary}
\affil[f]{C3S Electronics Development LLC., K{\"o}nyves K\'alm\'an krt. 12-14., Budapest, 1097, Hungary}
\affil[g]{Institute of Physics, E\"otv\"os University, P\'azm\'any P\'eter s\'et\'any 1/A, Budapest, 1117, Hungary}
\affil[h]{Department of Physics, Nagoya University, Furo-cho, Chikusa-ku, Nagoya, Aichi, Japan}
\affil[i]{The Hakubi Center for Advanced Research and Department of Astronomy, Kyoto University, Kyoto 606-8302, Japan}
\affil[j]{Department of Physics, University of Tokyo, 7-3-1 Hongo, Bunkyo, Tokyo 113-0033, Japan}
\affil[k]{Department of Physics, Rikkyo University, Nishi Ikebukuro 3-34-1, Toshimaku, Tokyo 171-8501, Japan}
\affil[l]{MTA-ELTE Astrophysics Research Group, P\'azm\'any P\'eter s\'et\'any 1/A, Budapest, 1117, Hungary}
\authorinfo{Further author information: (Send correspondence to N. Werner and J. {\v R}{\'i}pa)\\N. Werner: E-mail: wernernorbi@gmail.com\\  J. {\v R}{\'i}pa: E-mail: ripa.jakub@gmail.com}

\begin{document} 
\maketitle

\begin{abstract}
We propose a fleet of nanosatellites to perform an all-sky monitoring and timing based localisation of gamma-ray transients. The fleet of at least nine 3U cubesats shall be equipped with large and thin CsI(Tl) scintillator based soft gamma-ray detectors read out by multi-pixel photon counters. For bright short gamma-ray bursts (GRBs), by cross-correlating their light curves, the fleet shall be able to determine the time difference of the arriving GRB signal between the satellites and thus determine the source position with an accuracy of $\sim10^\prime$. This requirement demands precise time synchronization and accurate time stamping of the detected gamma-ray photons, which will be achieved by using on-board GPS receivers. Rapid follow up observations at other wavelengths require the capability for fast, nearly simultaneous downlink of data using a global inter-satellite communication network. In terms of all-sky coverage, the proposed fleet will outperform all GRB monitoring missions.
 
\end{abstract}


\section{INTRODUCTION}
\label{sec:intro}  

Gamma-ray bursts (GRBs) \cite{pir04,mesp06,ved09,kou12,ger12,gom12,ger13,kum15,wil17} have been studied for more than four decades, nonetheless their origin is still not fully understood. They are one of the most extreme explosive events ever observed. In spite of large efforts and numerous observations many open questions about their detailed physics remain. GRBs were discovered in 1967--73 by the Vela military satellites and were reported to the scientific community in the early seventies \cite{kle73,maz74}. They occur roughly once per day and are characterized by flashes of $\gamma$-rays typically lasting from a fraction of a second to thousands of seconds, in the energy range from several keV to few MeV. GRBs usually outshines any other gamma-ray source in the sky.

It has been found that the duration distribution of GRBs is bimodial \cite{kou93} which suggests that two distinct astrophysical populations are behind these energetic phenomena \cite{maz81b,kou93,bal03,bor04,mesa06,zha09}: the so called long GRBs (LGRBs) with a softer spectrum and with the prompt $\gamma$-ray emission lasting for more than 2 seconds were identified with the gravitational collapse of massive stars; and the so called short GRBs (SGRBs) with a harder spectrum and with the prompt $\gamma$-ray emission lasting for less than 2 seconds were proposed to originate in mergers of two compact objects such as neutron star - neutron star (NS-NS) or neutron star - black hole (NS-BH) \cite{ber14}.

The model that SGRBs originate in the merger of two NS has been recently confirmed by the detection of the gravitational wave (GW) signal GW170817 by the Laser Interferometer Gravitational-wave Observatory (LIGO) / Virgo collaboration\footnote{\url{https://www.ligo.org}}$^,$\footnote{\url{http://www.virgo-gw.eu}}. The detection of GW170817 was accompanied by a SGRB detected with the {\em Fermi}/Gamma-ray Burst Monitor (GBM) and the {\em INTEGRAL} satellite, followed by the observation of the GRB's afterglow and its host galaxy \cite{abb17a,abb17b,abb17c}. GW170817 confirmed the association of SGRBs with so called kilonovae \cite{tan13}.

Space observatories monitoring GRBs are expensive several hundred million USD missions, the result of which is that only a handful of them are launched per decade worldwide. Recently, however, affordable nanosatellites are starting to open up new opportunities for breakthrough science in space. We are proposing here to monitor and localize the GRBs, some of which are the electromagnetic counterparts of the gravitational wave signals, with a fleet of nanosatellites. The detection of GWs is among the most important breakthroughs of the decade in physics. Regular detections of electromagnetic counterparts to GW sources are paramount for the better understanding of these important phenomena. Our proposed mission, {\em Cubesats Applied for MEasuring and LOcalising Transients (CAMELOT)}, will enable all-sky monitoring and fast localization of GRBs, thus providing key observational data on these exciting phenomena.

After the upgrades of LIGO, following the finished O2 observing run, the new O3 run is scheduled to start in early 2019 and according to the current plans, the designed sensitivity will be achieved in 2021. The Kamioka Gravitational Wave Detector (KAGRA)\footnote{\url{http://gwcenter.icrr.u-tokyo.ac.jp/en}} is expected to start the first baseline cryogenic operation in 2019, and the observing runs with a full interferometer are expected in 2020s \cite{kag17}. Therefore, the combination of the all-sky coverage and the relatively precise localization provided by {\em CAMELOT} will enable a regular detection of electromagnetic counterparts of GW sources in the next decade, providing a critical contribution to our understanding of these exciting events.

\begin{figure}
\begin{center}
\includegraphics[width=0.70\linewidth]{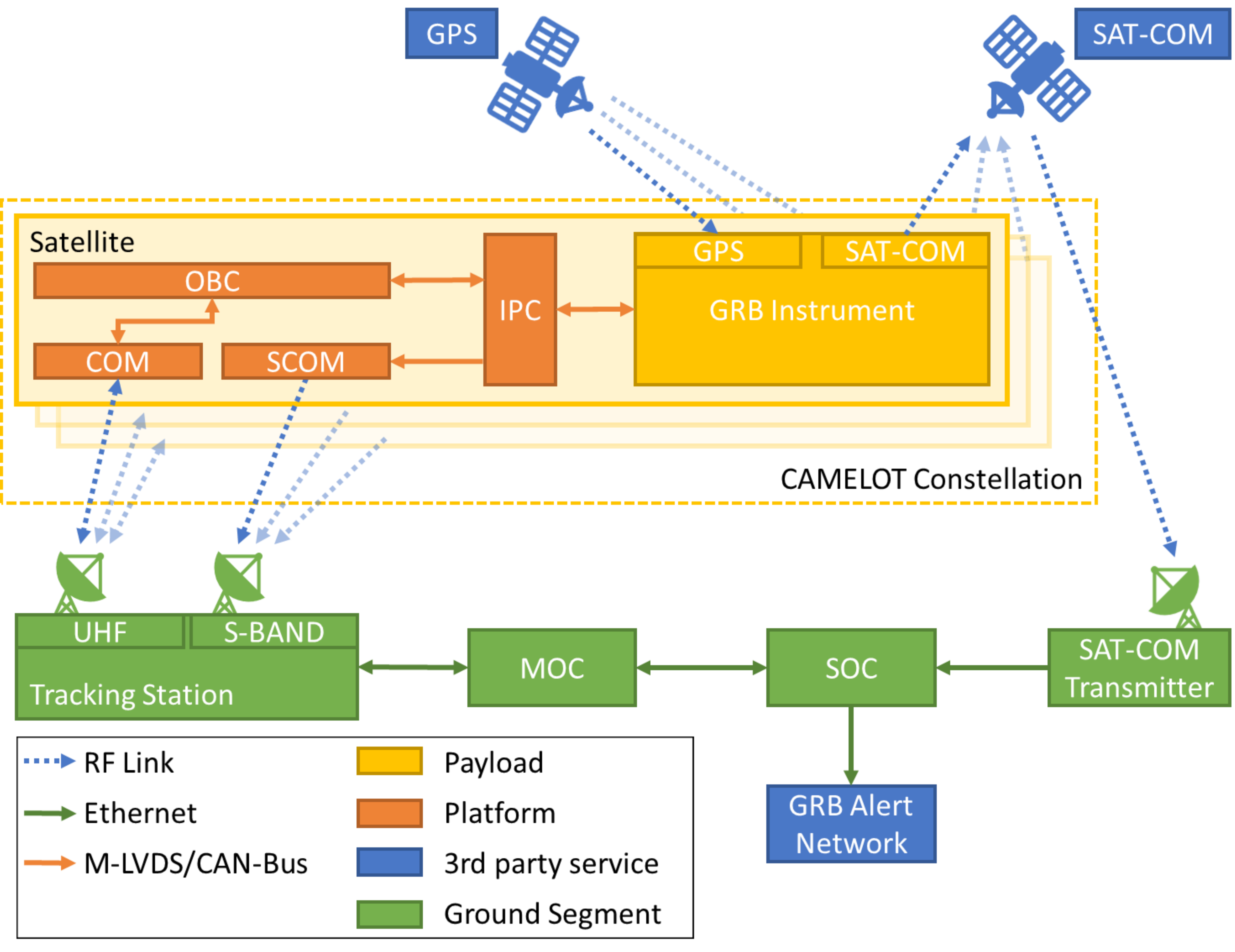}
\end{center}
\caption[]
{The data flow for the {\it CAMELOT} constellation. \label{fig:data_flow}}
\end{figure}

\section{Mission design}
\label{sec:platform_design}
The proposed constellation of nanosatellites shall provide all-sky coverage with high sensitivity and localization accuracy, as well as rapid data downlink following triggers. 

\subsection{The constellation operation concept}
According to our operation concept, the Mission Operation Center (MOC) will control the satellites by sending the operation tasks to the tracking station, which will upload the telecommands during the satellite passes. This means that, all scientific operations will be reviewed by the MOC. The tracking station will be able to operate and follow the satellites during the passes autonomously. Following an on-board GRB detection trigger, the satellite payload will downlink the data required for the localization to the Science Operation Center (SOC) using a global satellite communication module (we will use satellite communication because the alert has to be sent as soon as possible). After the localization of the GRB, the SOC will send a message to the GRB Alert Network. This concept can be seen in Figure~\ref{fig:data_flow}.

\subsection{The satellite platform}
\begin{figure}
\begin{center}
\includegraphics[width=0.46\linewidth]{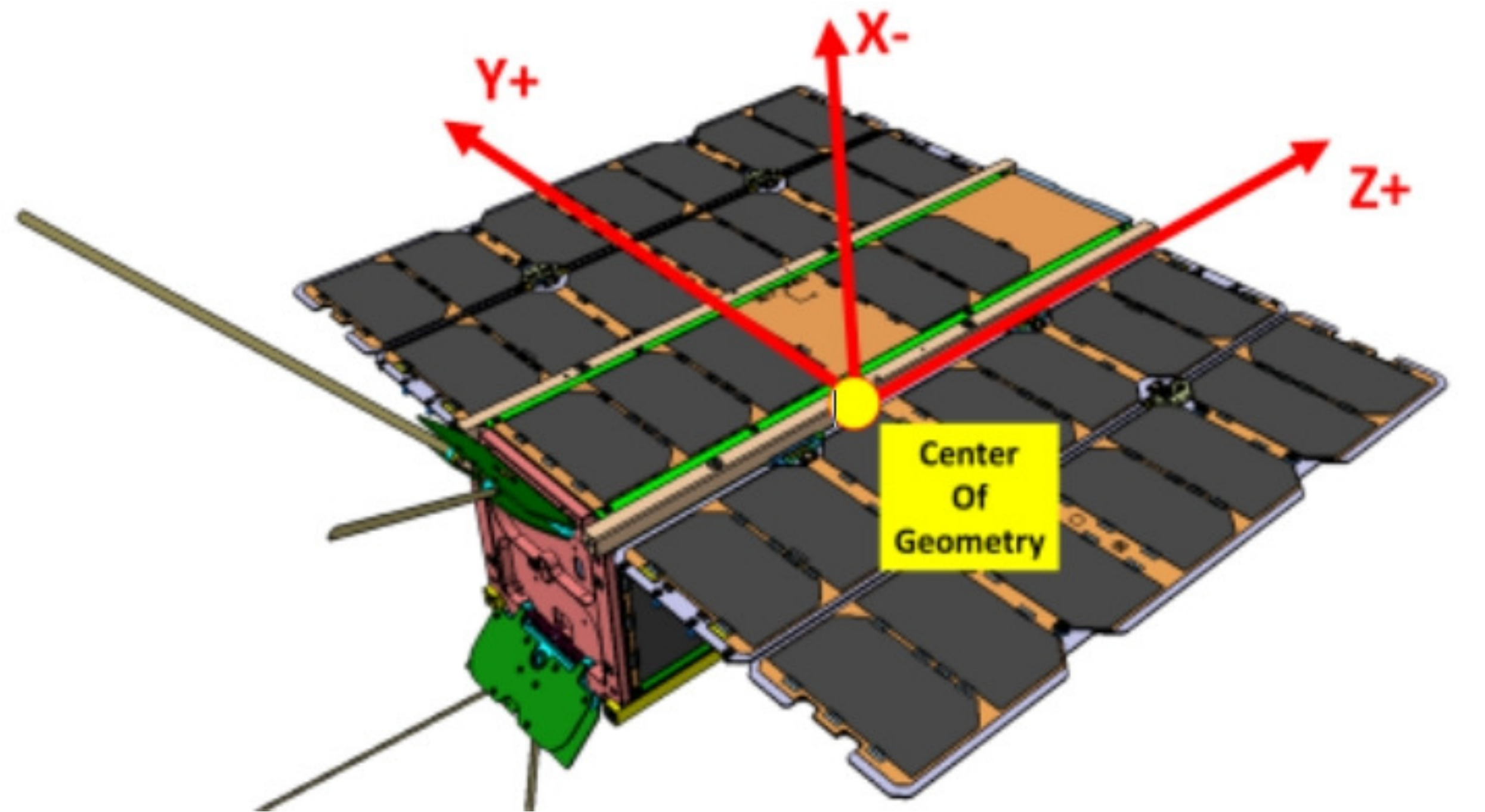}
\includegraphics[width=0.42\linewidth]{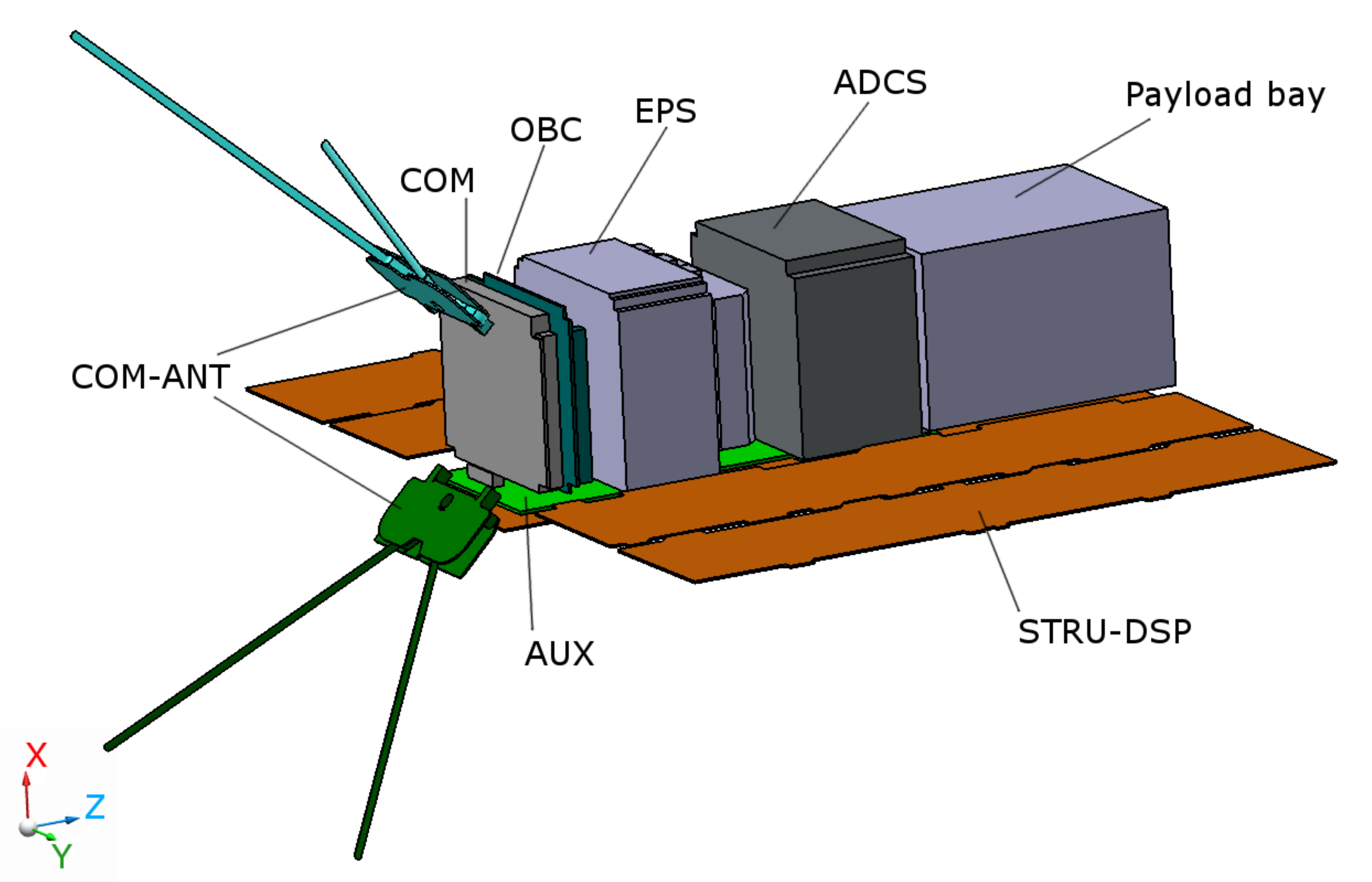}
\end{center}
\caption[example]
{ \label{fig:camelot}
A schematic view of the platform being developed for the {\em RadCube} satellite by C3S LLC. The same platform, with modifications, will be used for the proposed {\em CAMELOT} mission. The right panel shows the subsystems of the satellite, including the Payload Bay, Structure of Deployable Solar Panels (STRU-DSP), Communication Antenna (COM-ANT), auxiliary electronics (AUX), Communication UHF transceiver (COM), On-Board Computer (OBC), Electrical Power System (EPS), Attitude Determination and Control System (ADCS).}
\end{figure}

The satellite platform for the proposed {\em CAMELOT} constellation will be based on the 3U CubeSat developed for the RadCube mission by C3S LLC. The left panel of Fig.~\ref{fig:camelot} shows the cubesat and the right panel is a schematic view of its sub-systems. Table~\ref{tab:platform_spec} shows the specifications of the platform.

\begin{table}[h]
\centering
\begin{tabular}{ll}
\hline
\hline
Property                                   & Value                                                    \\
\hline
\\[-2.4ex]
Payload Mass                               & 2000\,g                                                  \\
Additional Payload Margin                  & 300\,g                                                   \\
Additional Platform Margin                 & 300\,g                                                   \\
\multirow{2}{*}{Available Payload Volumes} & $1 \times$ internal volume $100 \times 98 \times 98$\,mm \\
                                           & $2 \times$ external volumes $327 \times 83 \times 9$\,mm \\
Payload Orbit Average Power                & 5110\,mW (for sun-sync LEO)                              \\
Attitude Knowledge                         & $\sim 0.12^\circ$ ($3\sigma$)                            \\
Attitude Control                           & $<1.0^\circ$ ($3\sigma$)                                 \\
Design Orbit                               & 500--600\,km SSO                                         \\
Design Orbit Physical Lifetime             & 1--3 years (minimum)                                     \\
Mission Design Lifetime                    & 6 months                                                 \\
Downlink Frequency                         & UHF 400 MHz professional band                            \\
\multirow{4}{*}{Daily Average Downlink Data Rate} & 2.5\,Mbytes/24h at $>98$\,\% probability          \\
                                           & with single Hungarian ground station                     \\
                                           & additional 2.0\,Mbyte/24h reserved for                   \\
                                           & platform telemetry and system margin                     \\
\multirow{2}{*}{Platform Data Storage}     & 2 Gbytes for science data and telemetry                  \\
                                           & additional 2 Gbytes for platform and system margin       \\
Vibration and Shock Environment            & PSLV, FALCON 9, VEGA, ARIANE 5                           \\
Platform Design Radiation Environment      & 500--600\,km SSO equivalent 20 kRAD TID                  \\
\hline
\end{tabular}
\caption{The technical specifications of the 3U cubesat platform developed by C3S LLC.}
\label{tab:platform_spec}
\end{table}

\subsection{The detector concept}
\label{sec:payload_design}
To maximize the photon collecting area at $\sim100$ keV, where GRBs typically have a high flux, we plan to place two to four thin, and relatively large, $8.3\times15$~cm, CsI(Tl) scintillator detectors in an aluminum or carbon fiber reinforced plastics (CFRP) support structure on the surface of each satellite. The cubesat standard allows for lateral extensions of 9 mm, limiting the maximum thickness of the CsI(Tl) crystal to $\sim5-7$ mm, which is optimal for 100 keV photons and has a relatively low sensitivity above $\sim 300$ keV. The main advantage of CsI(Tl) scintillators is their large light output. Each scintillator will be read out by two Multi-Pixel Photon Counter (MPPC) Silicon Photomultipliers by Hamamatsu, which have a compact readout area and require a relatively low voltage. With this detector design, we are able to achieve a large photon collecting area in the $\sim10-300$ keV band, while leaving enough room for electronics in the payload bay of the cubesat. By using a multi-channel coincidence readout technique with two MPPCs, we add redundancy to our design, improve the light yield and reduce the noise. The time synchronization between the cubesats will be achieved by using an on-board GPS receiver and the detected $\gamma$-ray photons will be timestamped with an accuracy of $<100\mu$s. For a detailed description of the payload design, the GPS timestamping unit and the gamma-ray detectors see the other two articles submitted to this SPIE conference by the {\em CAMELOT} collaboration (P\'al et al. and Ohno et al.).

\begin{figure}[h]
\begin{center}
\includegraphics[width=0.24\linewidth]{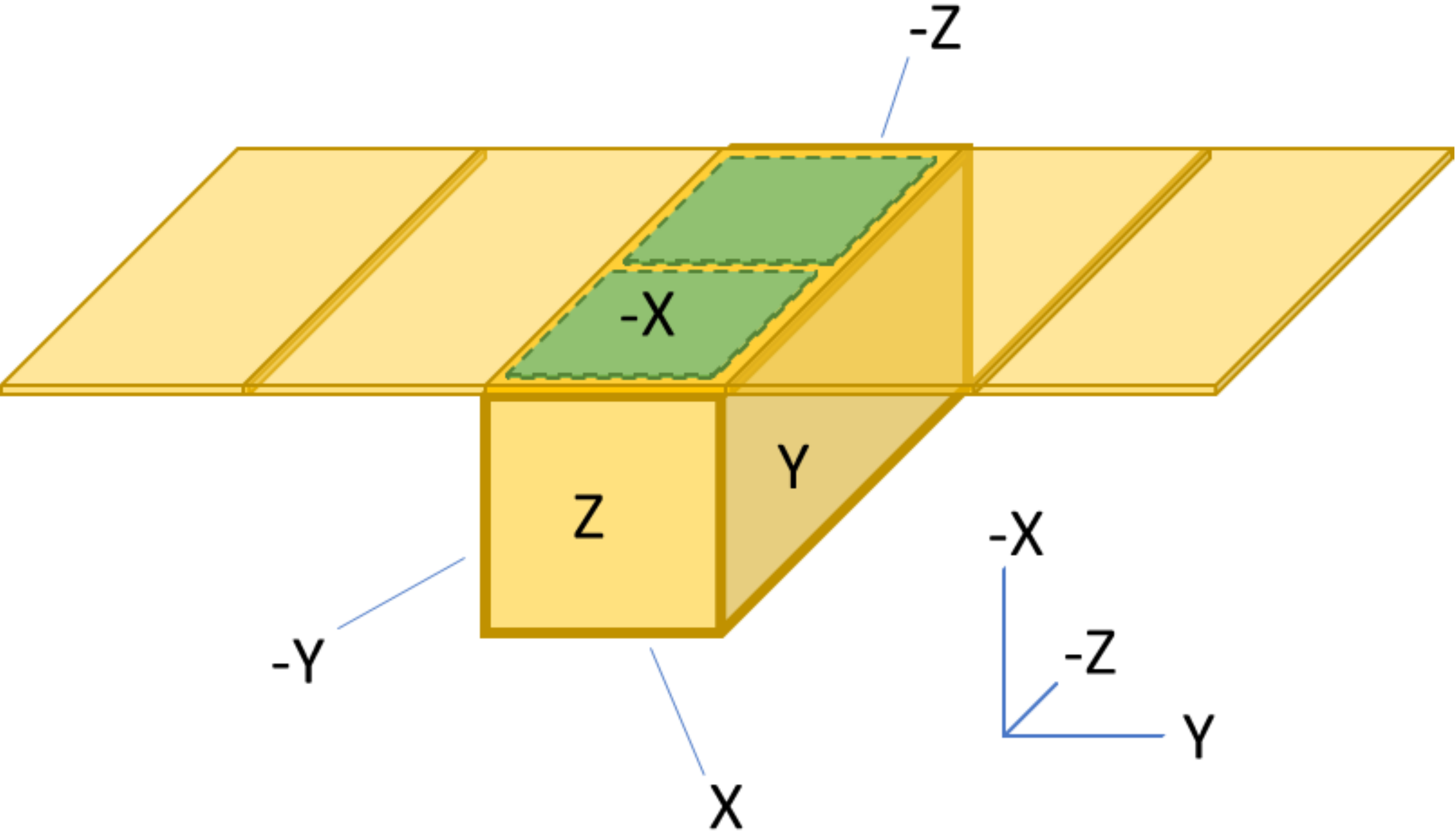}
\includegraphics[width=0.24\linewidth]{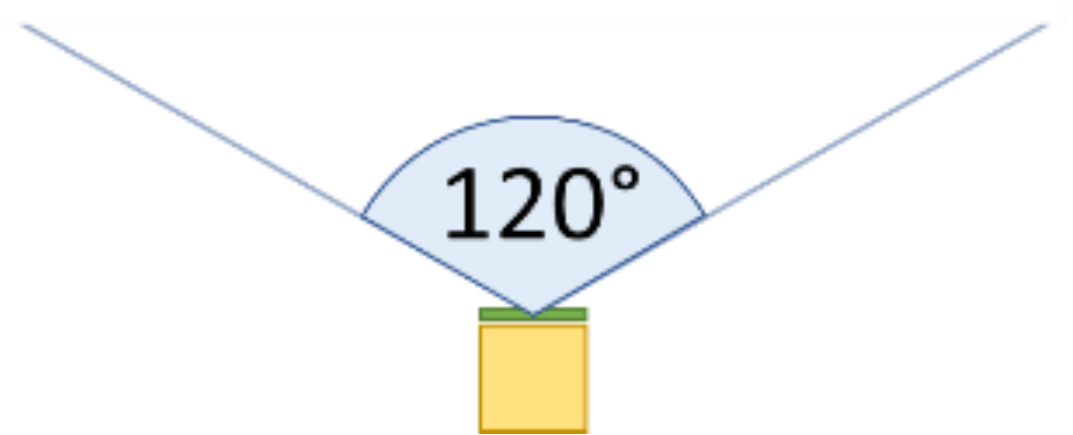}
\includegraphics[width=0.24\linewidth]{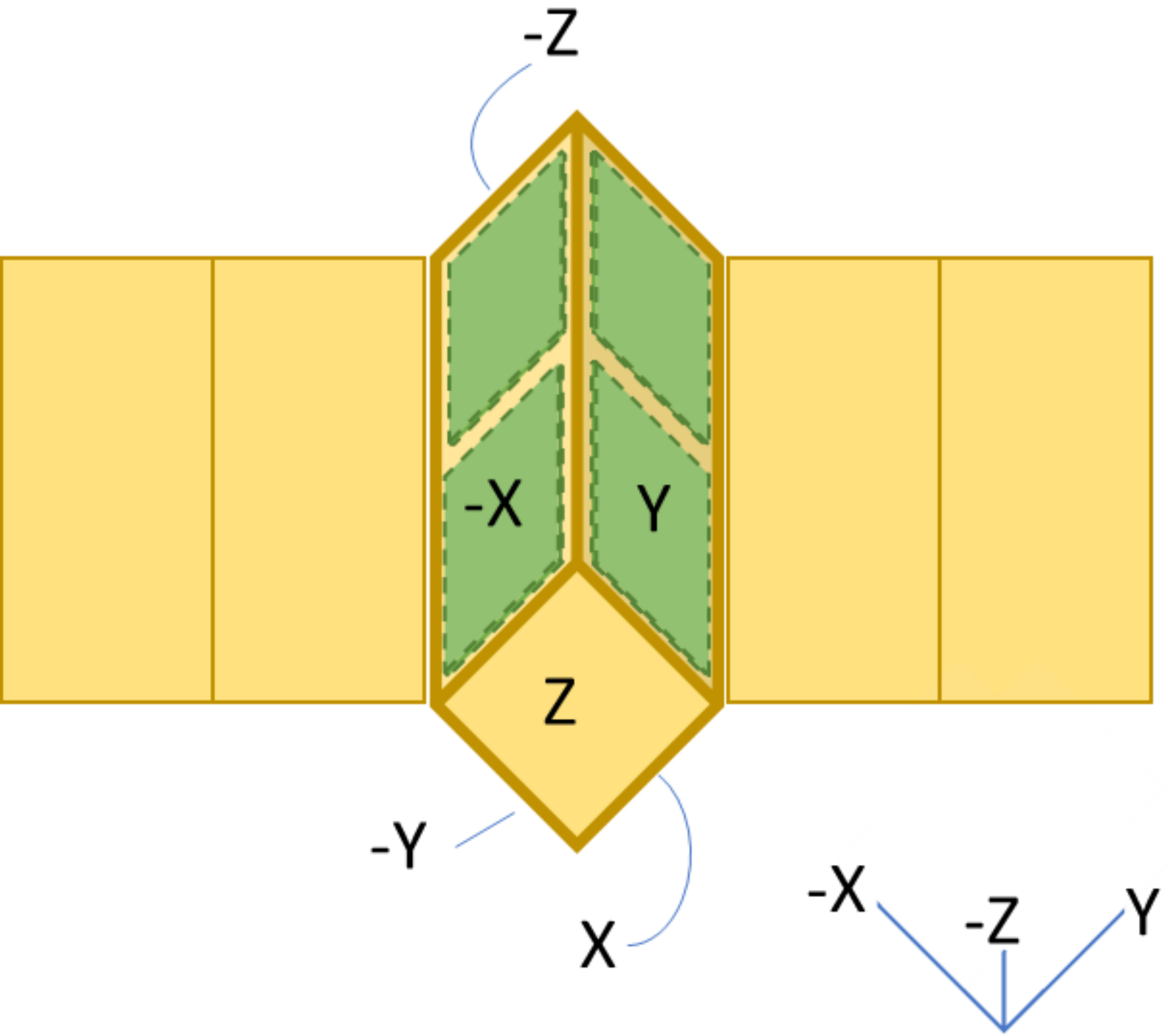}
\includegraphics[width=0.24\linewidth]{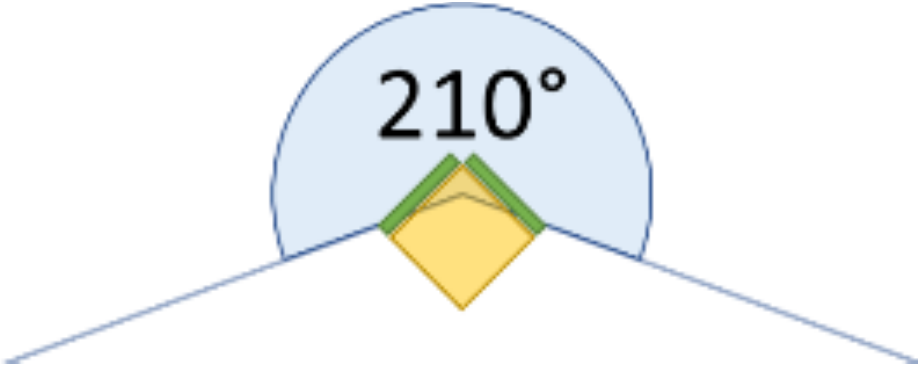}
\end{center}
\caption[example]
{ \label{fig:det_conf_fov_one_side}
Possible one-sided (left two panels) and two-sided (right panels) detector configurations and their effective fields of view (see the text for more detail).}
\end{figure}

The detectors will be placed on one or two of the long sides of the satellite (see Fig.~\ref{fig:det_conf_fov_one_side}). In our visibility calculations in the next section, we make the conservative assumption that if we only place the detectors on one side of the satellite, the effective field of view (FoV) will be approximately a cone with an apex angle of $120^\circ$ (see the left two panels of Figure~\ref{fig:det_conf_fov_one_side}). Correspondingly, if the detectors are placed on two perpendicular sides of the platform, then our effective FoV will significantly increase to $210^\circ$ along the X, Y plane (see the right two panels of Figure~\ref{fig:det_conf_fov_one_side}). However, our actual field of view will be larger than these simple assumptions. Our GEANT4 simulations show that about 60\% of the gamma ray photons arriving at a satellite from a GRB at the opposite side and penetrating through the whole cubesat platform, will be absorbed or scattered. 40\% of the photons will still arrive at the detectors. The default attitude of the satellites will be such that the detectors will be aligned with the zenith, maximizing the unobstructed sky coverage. 

\begin{figure}
\begin{center}
\includegraphics[width=0.48\linewidth]{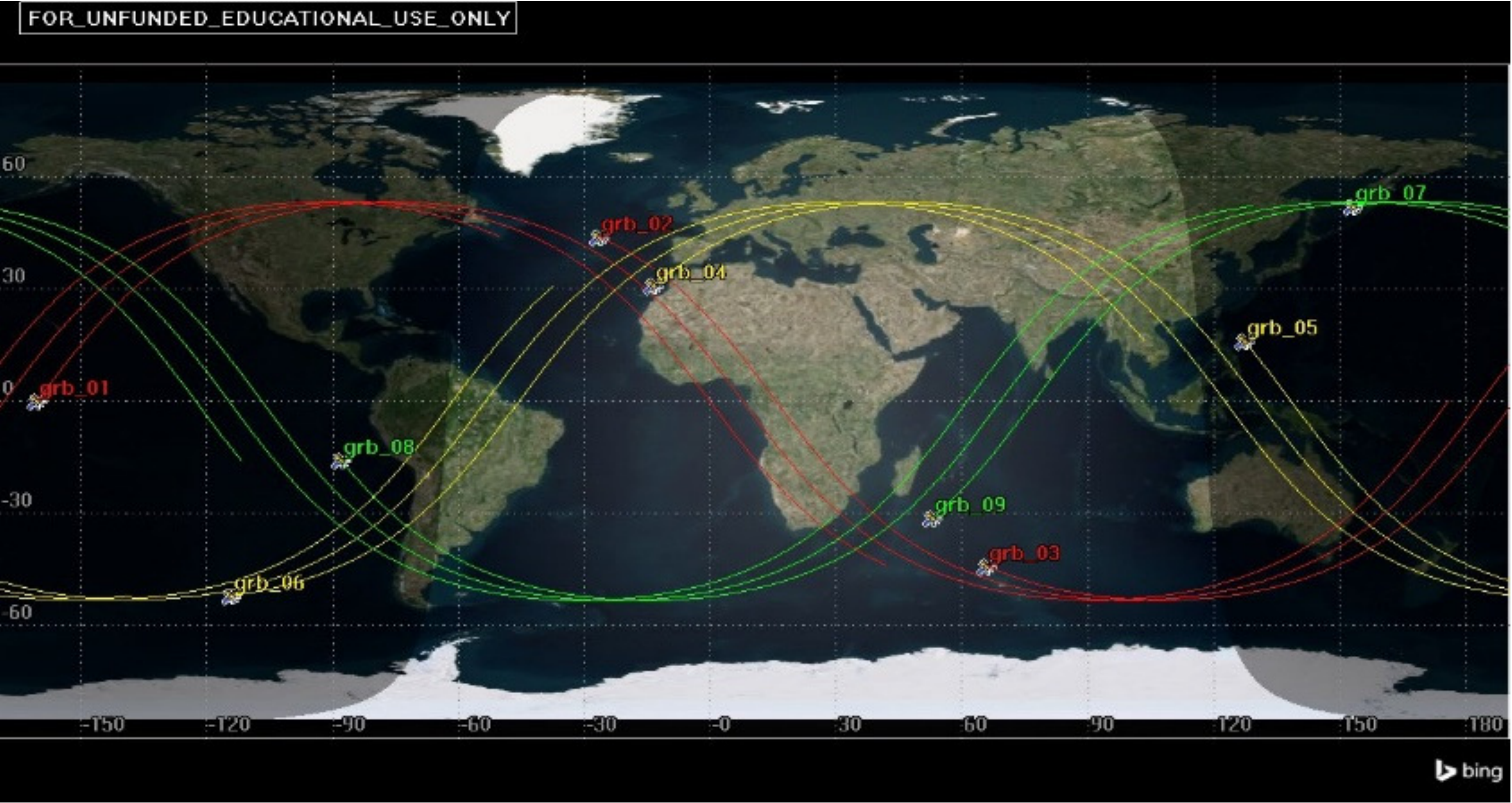}
\includegraphics[width=0.50\linewidth]{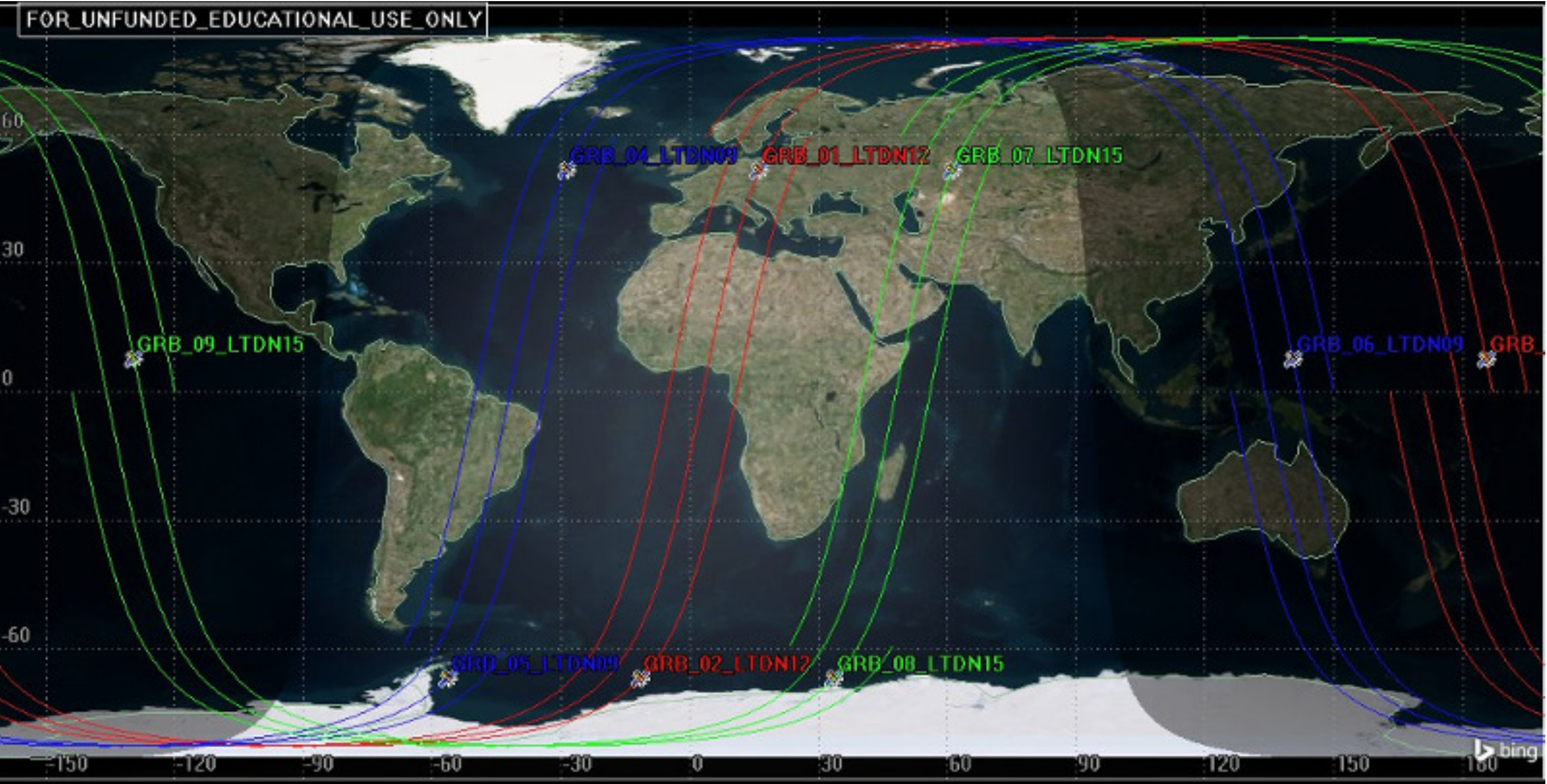}
\end{center}
\caption[example]
{ \label{fig:orbits}
Nine satellites on low-Earth orbits (altitude of 550 km) in three orbital planes with inclination of $53^\circ$ ({\it left}) or on sun-synchronous orbits with an inclination of $97.6^\circ$ ({\it right}) could, in principle, provide both all-sky coverage and the required localization accuracy.}
\end{figure}

\subsection{Orbits and sky visibility}

We study two different sets of orbits: the so called {\it Walker low-Earth orbits} (LEO; altitude $\sim550$~km) with an inclination of $53^\circ$ and {\it Sun-synchronous orbits} (SSO) with an inclination of $97.6^\circ$ (also LEO). We determine the sky coverage for two different detector configurations: with a set of detectors covering one side of the satellites and with two perpendicular sets of detectors covering two sides of the satellites. The detector configurations are described in Section~\ref{sec:payload_design} and in Figure~\ref{fig:det_conf_fov_one_side}. The analysis is performed using the AGI STK software package (for educational use) for nominal operations, assuming that the detectors are aligned with the zenith. The requirement driving the design of the constellation is to continuously cover most of the sky by at least three satellites in order to be able to localize transients using the triangulation method.

\begin{figure}[t]
\begin{center}
\includegraphics[width=0.49\linewidth]{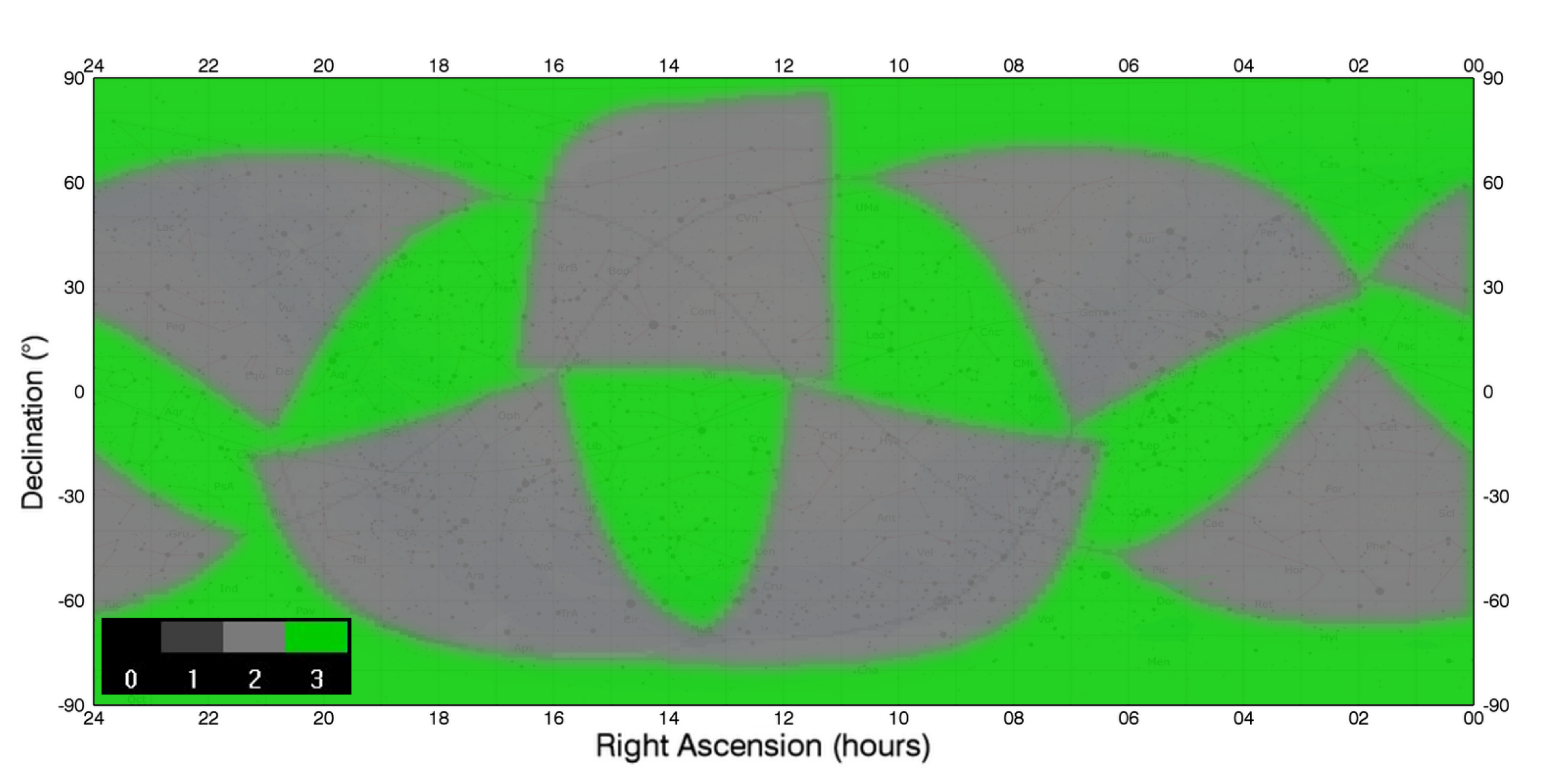}
\includegraphics[width=0.49\linewidth]{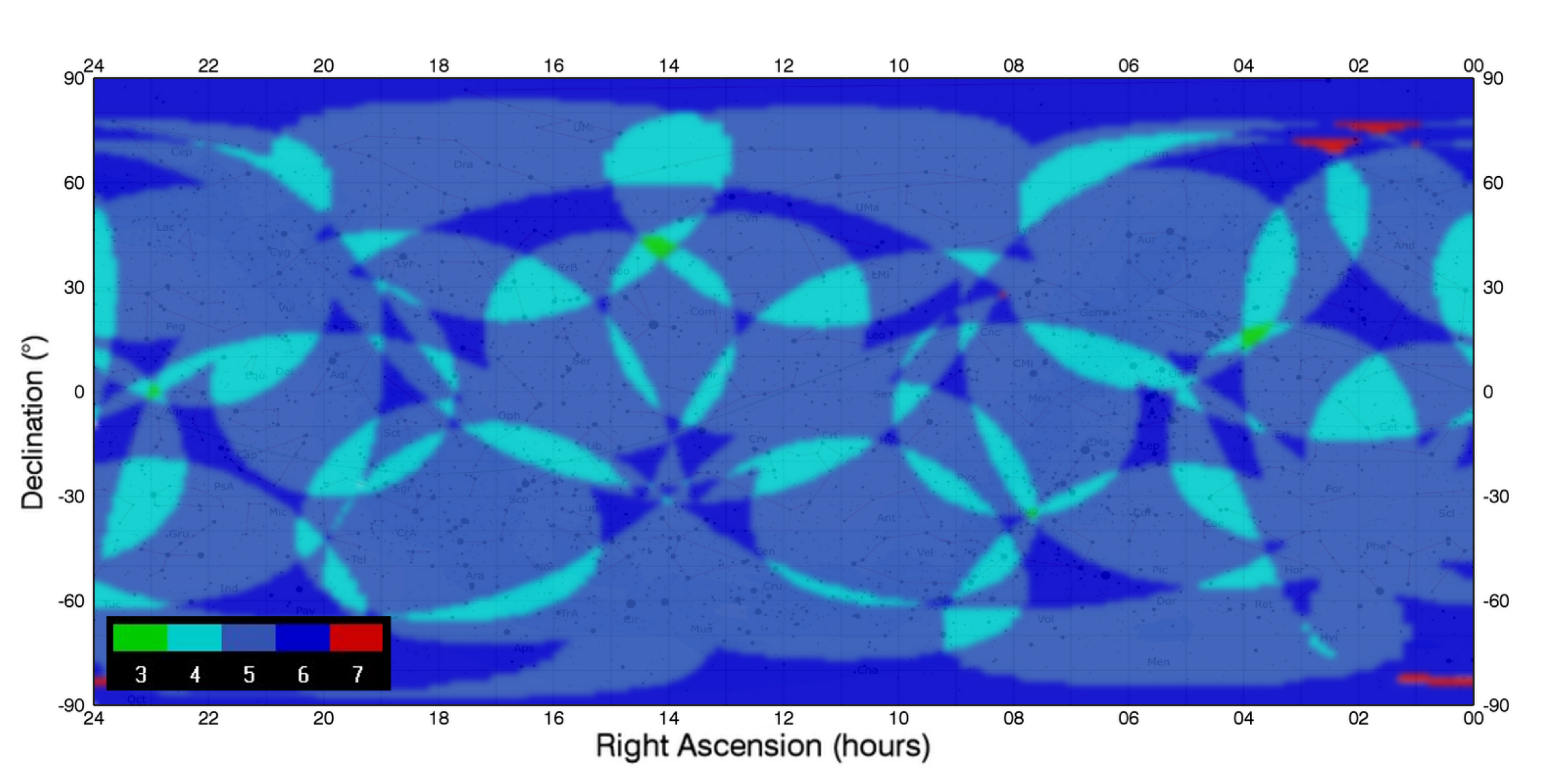}
\end{center}
\caption[example]
{ \label{fig:sky_coverage_walker}
Sky coverage for a constellation of 9 satellites in low-
Earth (550 km) Walker orbits with an inclination of 53$^{\circ}$ for a set of detectors only covering one side of each cubesat ({\it left}) and for two perpendicular sets of detectors ({\it right}). The color indicates the number of satellites simultaneously covering a given part of the sky. See the legend in the lower left corner of the figures for the color corresponding to the given number of satellites.}
\end{figure}

\begin{figure}[t]
\begin{center}
\includegraphics[width=0.48\linewidth]{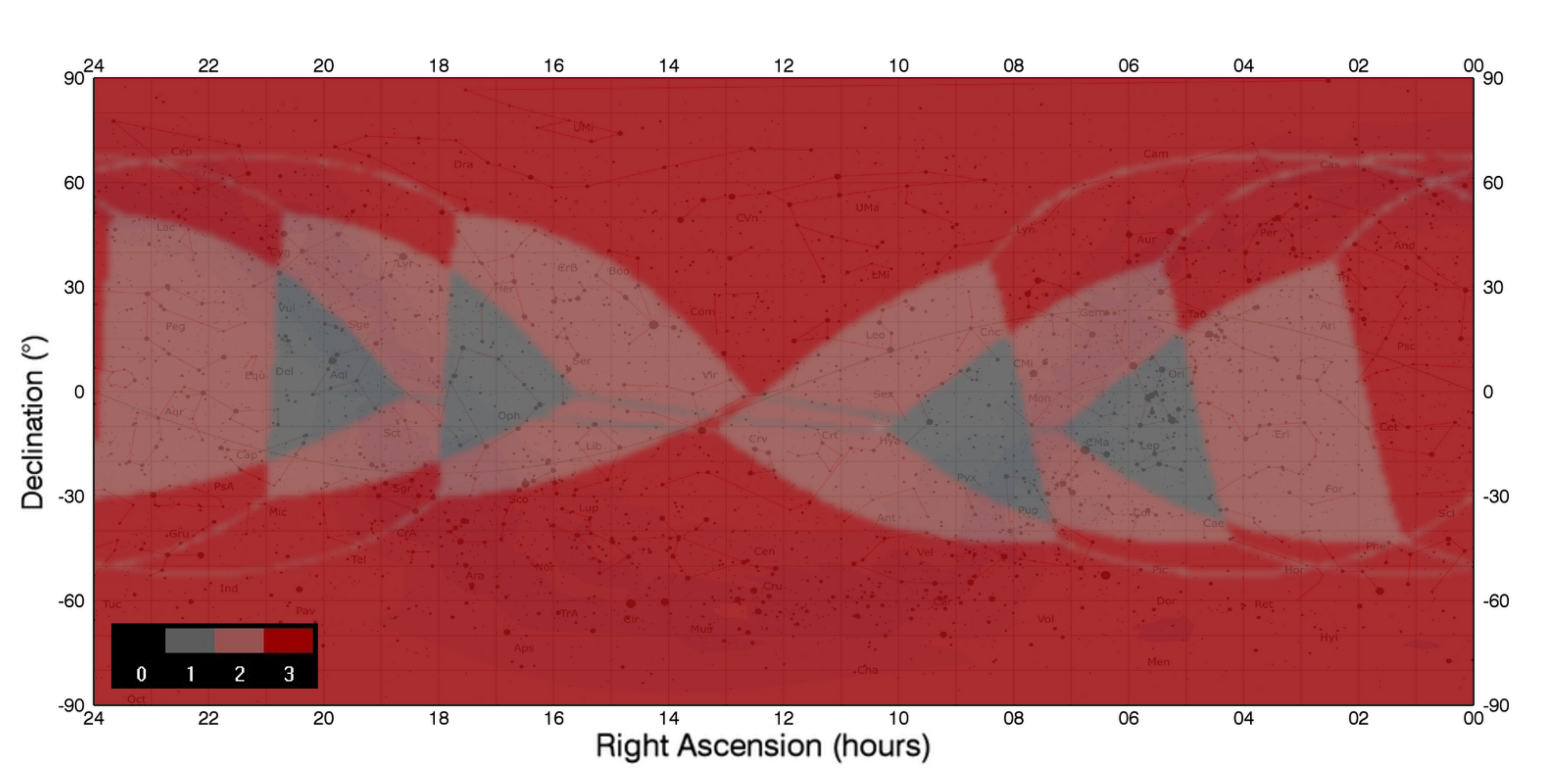}
\includegraphics[width=0.48\linewidth]{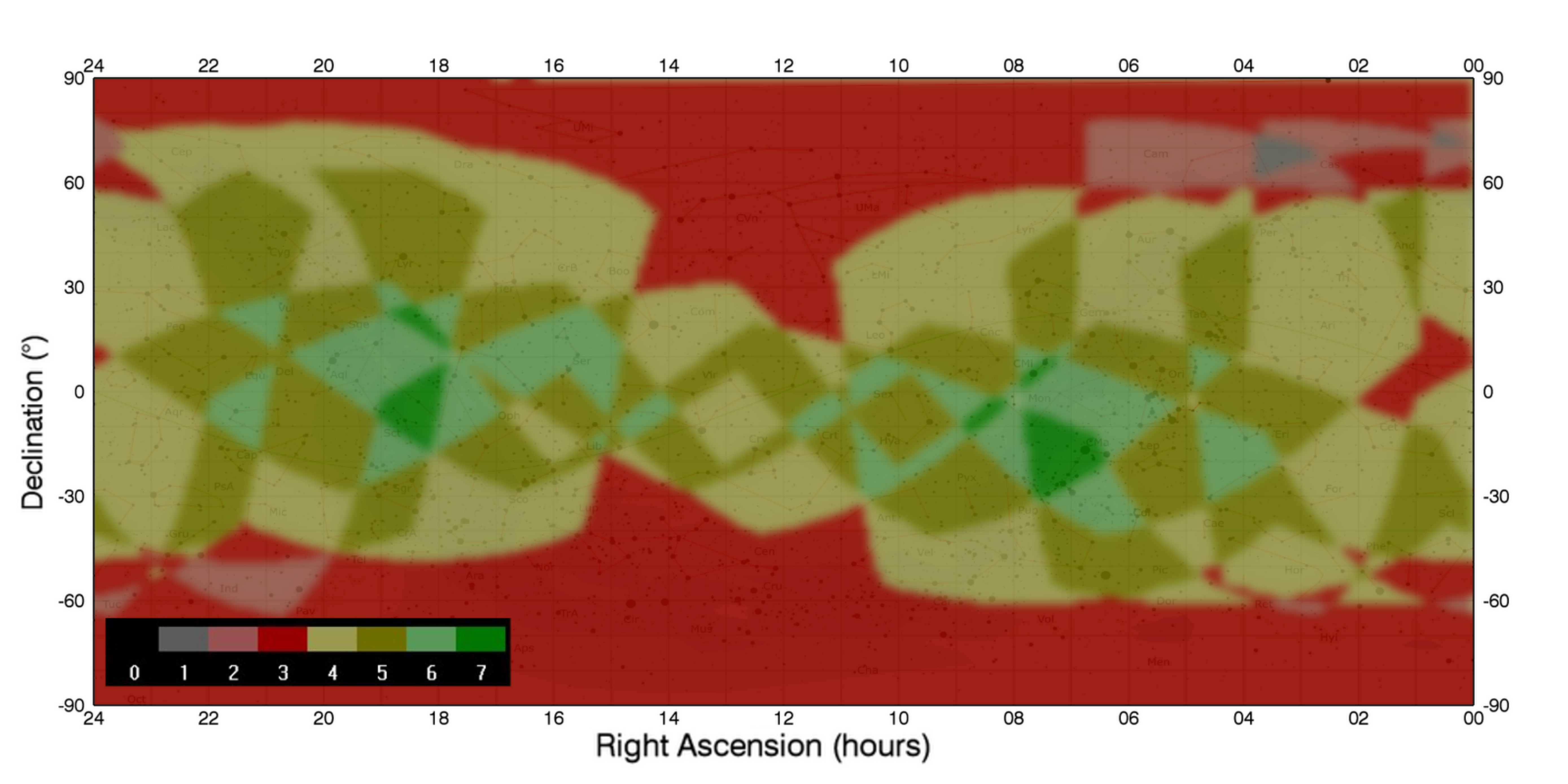}
\end{center}
\caption[example]
{ \label{fig:sky_coverage_polar}
Sky coverage for a constellation of 9 satellites in low-
Earth (550 km) polar SSO with a set of detectors only covering one side of each cubesat ({\it left}) and for two perpendicular sets of detectors ({\it right}). The color indicates the number of satellites simultaneously covering a given part of the sky. See the legend in the lower left corner of the figures for the color corresponding to the given number of satellites.}
\end{figure}

Figure~\ref{fig:sky_coverage_walker} and Figure~\ref{fig:sky_coverage_polar} show the covered sky area for a constellation of 9 satellites in Walker and SSO orbits, respectively. Table~\ref{tab:sky_coverage_9sat} shows the percentage of the sky simultaneously covered by a given number of satellites for a constellation of 9 satellites on Walker and SSO orbits with the two different detector configurations. We find that nine satellites with two perpendicular sets of detectors in three orbital planes (see Figure~\ref{fig:orbits}), with three satellites in each plane, could in principle provide both all-sky coverage and the required localization accuracy.

While satellites with both detector configurations would provide the required all-sky coverage, a continuous nearly all-sky coverage with at least three detectors at all times, required for the localization by triangulation, is provided only by satellites with two perpendicular detectors. The Walker orbits with inclination of $53^\circ$ provide a significantly better coverage, covering 95.3\% of the sky with five satellites at all times. Unfortunately, launch opportunities to such orbits are rare on the launch market. However, launch opportunities on small rockets dedicated to nanosatellites are expected to become much more common in the next few years.

\begin{table}[h]
\newcommand{\multirot}[1]{\multirow{4}{*}[-2.0ex]{\rotcell{\rlap{#1}}}}
\centering
\begin{tabular}{c|cccccccccc}
\hline
\hline
\\[-2.4ex]
                       & \multicolumn{8}{c}{Simultaneous sky coverage by at least \# satellites}         &   &   \\
Constellation/detector configuration &  1    &  2    &  3    &  4    &  5   &  6    &  7    &  8    &   &   \\
\hline
\\[-1.7ex]
Walker/one set           & 100.0 & 99.0  & 40.7  &  -    &  -    &  -    &  -   &  -    & \multirow{4}{*}{\multirot{Covered Sky}} & \multirow{4}{*}{\multirot{~~~~Area [\%]}}  \\[0.7ex]
Walker/two perpendicular sets           & 100.0 & 100.0 & 100.0 & 99.6 & 95.3  & 53.0  & 15.3 & - &   &   \\[0.7ex]
SSO/one set              & 99.9  & 88.9  & 50.9  &  -    &  -    &  -    &  -   &  -    &   &   \\[0.7ex]
SSO/two perpendicular sets              & 99.9  & 99.5  & 97.8  & 75.6  & 41.2  & 14.0  &  -   &  -    &   &   \\
\hline
\end{tabular}
\vspace{0.3cm}
\caption{Percentage of the sky simultaneously covered by a given number of satellites for a fleet of 9 satellites. Two types of low Earth (altitude of 550 km) orbits are investigated: Walker orbits with inclination of $53^\circ$ and SSO with inclination of $97.6^\circ$. Two different detector configurations are considered: a set of detectors covering one side of the satellite and two perpendicular sets of detectors.}
\label{tab:sky_coverage_9sat}
\end{table}

\begin{figure}[h]
\begin{center}
\includegraphics[width=0.7\linewidth]{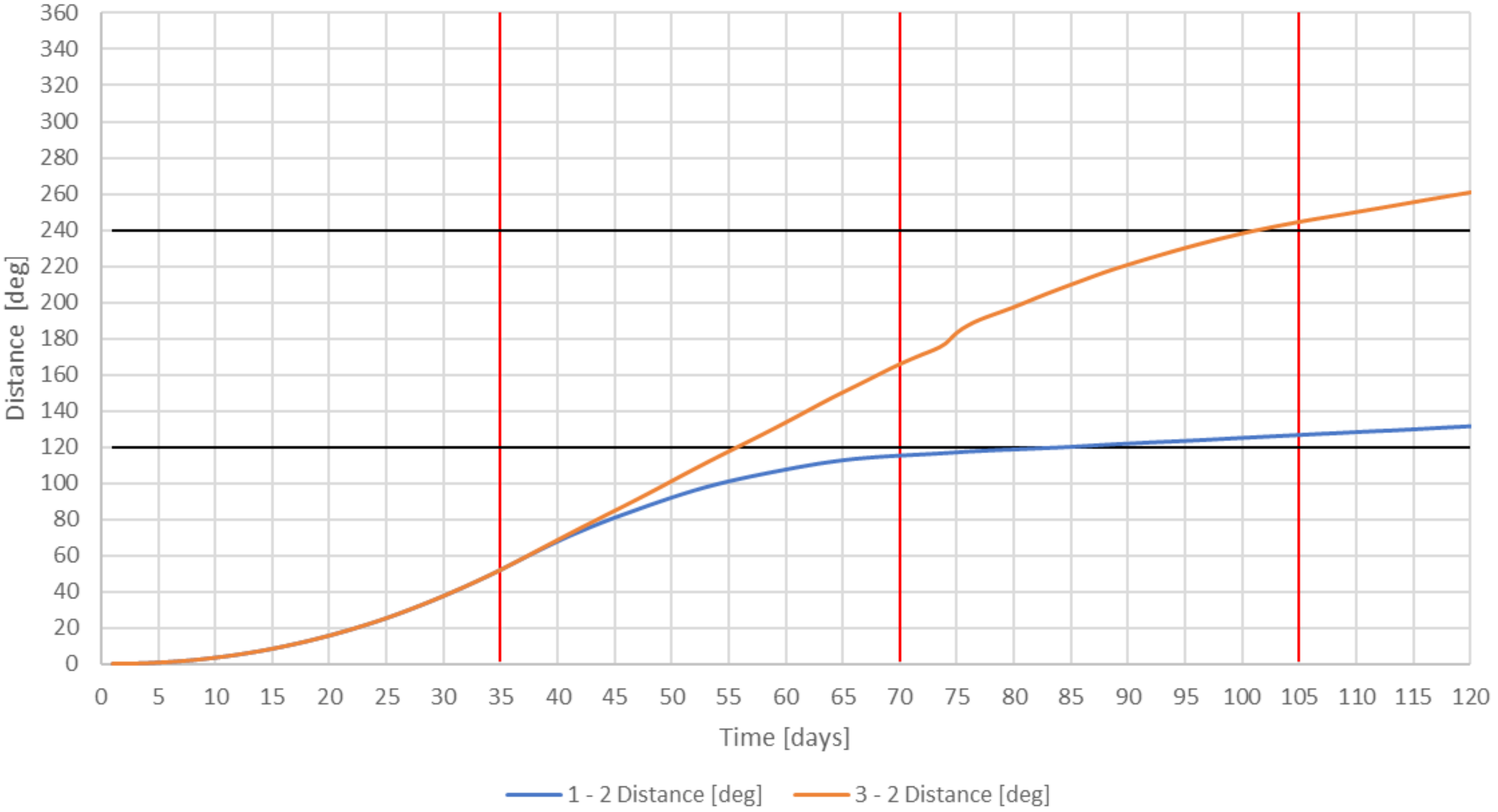}
\end{center}
\caption[example]
{ \label{fig:sat_separ}
Following a common launch into a circular orbit at an altitude of 550 km, three 3U cubesats can be evenly distributed along the orbit by a sequence of maneuvers changing the atmospheric drag using the attitude control, in about 100 days.}
\end{figure}

The deployment of the full 9-satellite constellation will require three three launches into three different orbital planes. We performed an analysis to determine a sequence of maneuvers to distribute the satellites uniformly along the orbit, after a common launch to a given orbital plane, using only attitude control and atmospheric drag. By changing the attitude to such that the satellite experiences maximum atmospheric drag its altitude will decrease, resulting in a shorter orbital period. The distance between two satellites with different attitudes and drags will thus increase. The goal is to have the satellites  $120^\circ$ from each other along the orbit. The sequence of maneuvers includes setting the drag to maximum for one satellite and to minimum for the other cubesat. About half of the duration of the separation maneuver is spent in low-drag and the other half in high-drag configuration by all satellites, so that all cubesats will have the same final orbital altitude and period. The distance between the satellites versus time for our simulation is plotted in Figure~\ref{fig:sat_separ}.  
We find that the satellites can be distributed uniformly along the orbit in about 100 days. Maneuvers involving changes in drag can also be used to correct for perturbations which might otherwise disrupt the optimal separation of satellites.

The preliminary calculation of the in-orbit lifetime of the satellites due to the decay by atmospheric drag was performed using the DRAMA 2.0 software package by ESA. The ``ECSS Sample Solar Cycle'' was used to predict the solar and geomagnetic activity. Table~\ref{tab:lifetime_orbit_decay} shows the lifetimes for various altitudes. To provide both sufficient lifetime for the mission and comply with the requirements for space debris mitigation, the ideal altitude range for the satellites is between $\sim 500$\,km and 600\,km. 

\begin{table}[h]
\centering
\begin{tabular}{ccc}
\hline
\hline
\\[-2.4ex]
Altitude         & Semi major axes  & Lifetime \\
$[$km]           & [km]             & [years]  \\
\hline
\\[-2.4ex]
450             & 6821             & 2.43      \\
500             & 6871             & 3.65      \\
550             & 6921             & 5.92      \\
600             & 6971             & 16.34     \\
618.7           & 6989.7           & 24.43     \\
650             & 7021             & 32.24     \\
\hline
\\[-2.4ex]
\end{tabular}
\caption{Lifetime versus initial altitude of the satellites.}
\label{tab:lifetime_orbit_decay}
\end{table}

\subsection{The expected particle background}
\label{background_estimation}

The sensitivity of our detectors will be significantly affected by the radiation environment. In particular, if we place our constellation to polar SSO orbits, we will experience increased background count rates around the polar regions due to the high flux of electrons. Therefore, a study of the expected particle background is especially important for designing the constellation and the trigger method. 

\begin{figure}[h]
\begin{center}
\includegraphics[width=0.49\linewidth]{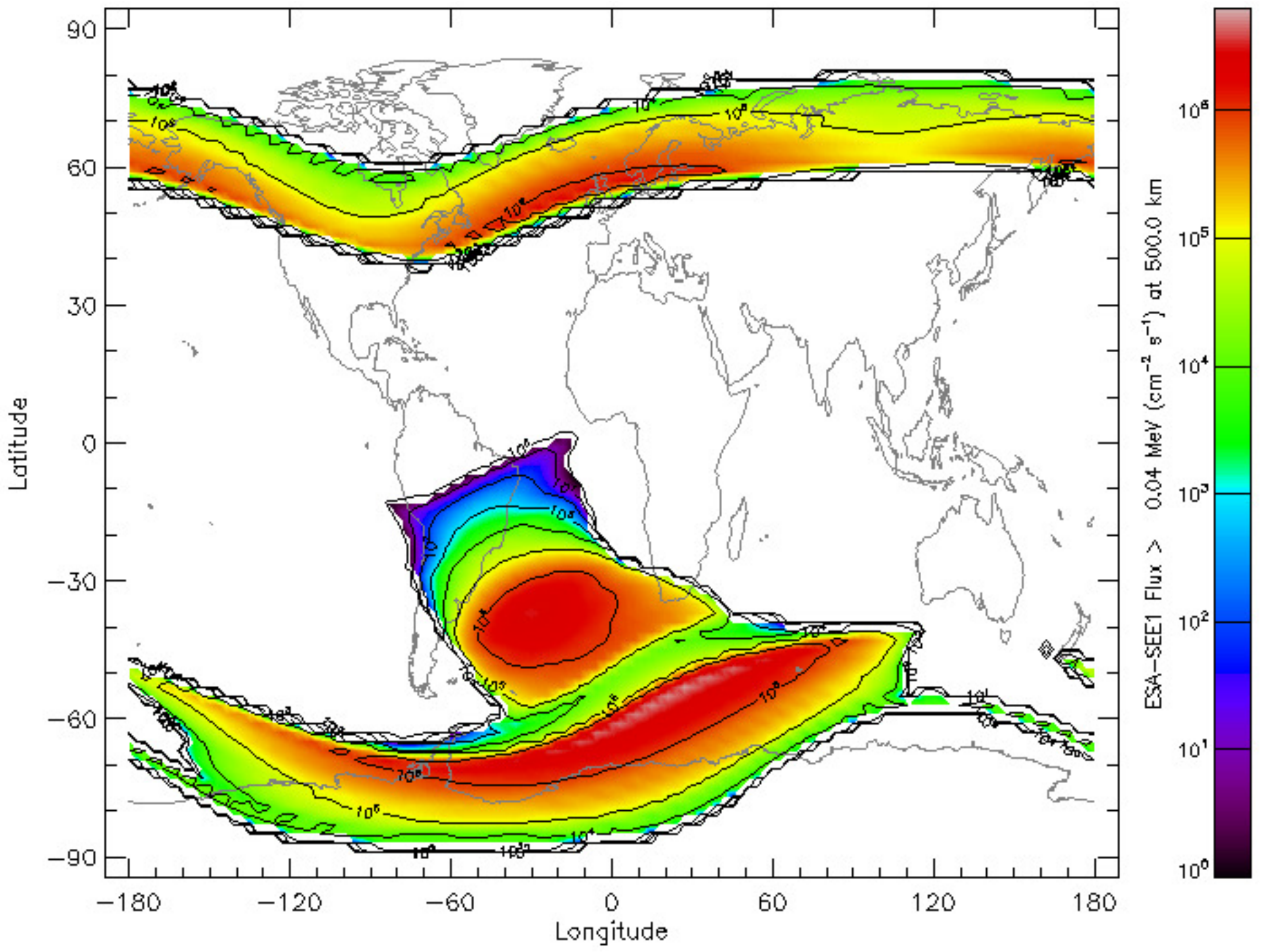}
\includegraphics[width=0.49\linewidth]{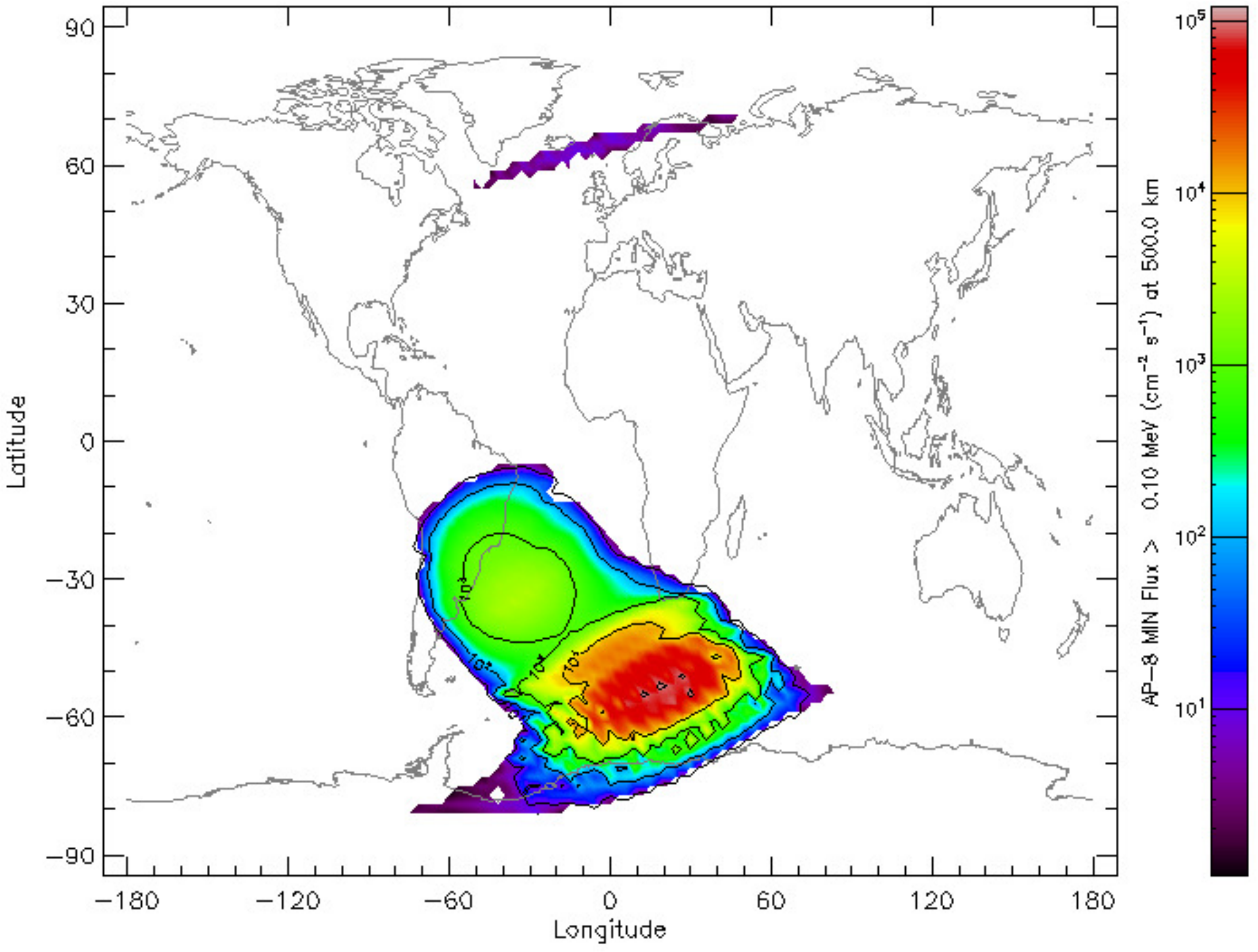}
\end{center}
\caption[example]
{ \label{fig:e_p_flux_map}
{\it Left:} Map of electron fluxes with energies $>40$\,keV at the altitude of 500\,km based on the ESA-SEE1 model for solar minimum (credit: SPENVIS and \cite{vam96}). {\it Right:} Map of proton fluxes with energies $>0.1$\,MeV at the altitude of 500\,km based on the AP-8 model. (credit: SPENVIS and \cite{saw76}).}
\end{figure}

The left panel of Figure~\ref{fig:e_p_flux_map} shows a flux map for electrons with energies higher than 40\,keV at the altitude 500\,km as obtained from the European Space Agency’s (ESA) SPace ENVironment Information System (SPENVIS)\footnote{\url{https://www.spenvis.oma.be}} developed by the Royal Belgian Institute for Space Aeronomy (BIRA-IASB). The map is based on the ESA-SEE1 model, which is an update of NASA's AE-8 model employing the {\em CRRES}/MEA satellite data from 1990--1991 \cite{vet91,vam96} for the solar minimum. The right panel of Fig. 10 shows a map of proton fluxes with energies higher than 0.1\,MeV at the same altitude (also obtained from SPENVIS). The map is based on NASA's AP-8 model \cite{saw76} for the solar minimum.

\begin{figure}[h]
\begin{center}
\includegraphics[width=0.55\linewidth]{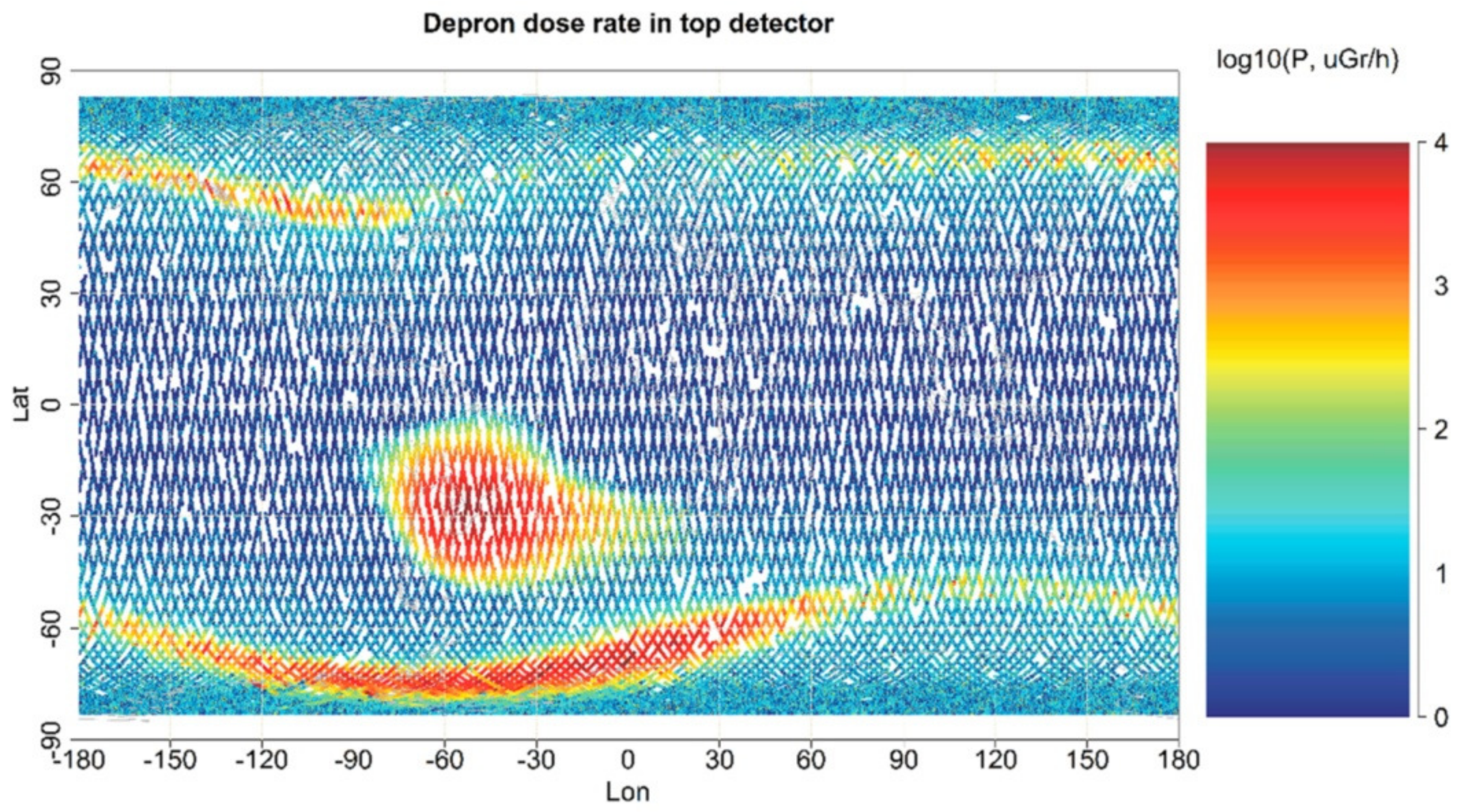}
\includegraphics[width=0.44\linewidth]{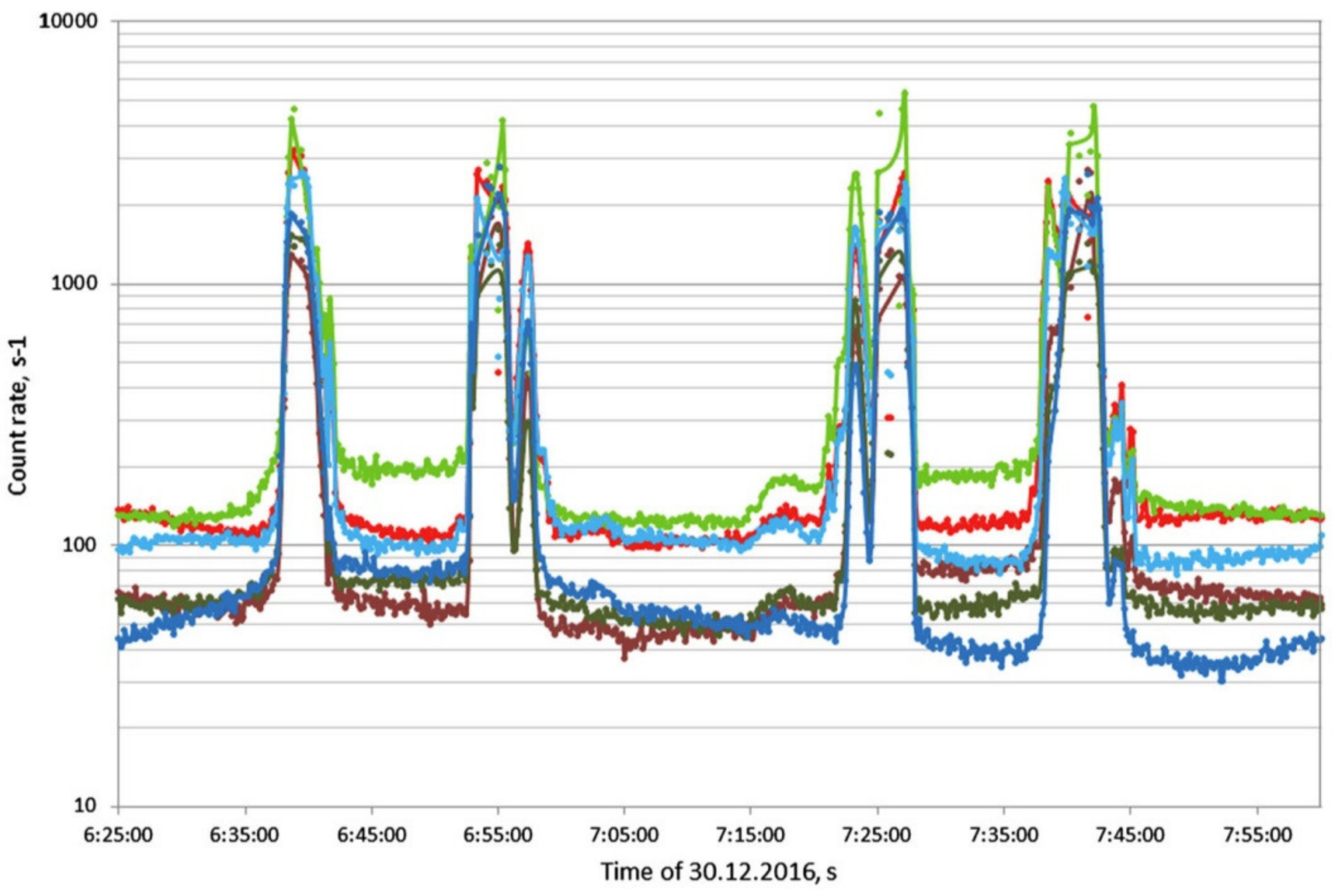}
\end{center}
\caption[example]
{\label{fig:fluxes_lomonosov}
{\it Left:} Map of the radiation environment for SSO as measured by the {\it Lomonosov}/DEPRON instrument \cite{ben18}. {\it Right:} Count rate variations in the {\it Lomonosov}/BDRG NaI(Tl) scintillator on polar SSO: BDRG-1 20--35\,keV (light red line), 6--10\,keV (dark red line), BDRG-2 20--35\,keV (light green line), 6--10\,keV (dark green line), BDRG-3 20--35\,keV (light blue line), 6--10\,keV (dark blue line)\cite{sve18}.}
\end{figure}

The left panel of Figure~\ref{fig:fluxes_lomonosov} shows a map of the radiation dose rate of charged particles (electrons and protons) at LEO as measured by the {\em Lomonosov}/Depron instrument. The {\em Lomonosov} satellite was launched to a low-Earth (altitude of $\sim 480$\,km) sun-synchronous polar orbit ($i=97.2^\circ$) \cite{sad17} with charged particle and gamma-ray detectors on-board. Therefore, the data collected by this mission provide a good reference for the expected background for any soft gamma-ray detector on high-inclination LEO. One of the instruments on board the {\em Lomonosov} satellite is the BDRG instrument, dedicated to detecting GRBs. It consists of three modules with NaI(Tl) and CsI(Tl) scintillators read out by Hammamatsu R877 photomultiplier tubes (PMT). The scintillators are enclosed by a thin layer of aluminum and they are sensitive to gamma-rays from 10\,keV to 3.0\,MeV \cite{sve18}. The right panel of Figure~\ref{fig:fluxes_lomonosov} shows the count rate variations in the NaI(Tl) scintillator when crossing the polar regions.

According to the {\em Lomonosov}/BDRG observations, the count rate in the 10--450\,keV energy range increases $\sim 60$ times in the polar regions compared to the equator. Such high background count rates, which last for about 1/4 to 1/3 of the duration of the polar orbit, make the detection of GRBs difficult. The fast background variations can also cause false triggers, which can be avoided by disabling the algorithm during the polar passes.

If we place the constellation to SSO we will most likely loose up to 30--40\,\% of observing time per satellite due to high background and its fast variations in the polar regions. Additional time will be lost during the passes through the South Atlantic Anomaly (SAA). If the constellation is launched to SSO then we will most likely require more than 9 satellites to compensate for the loss of observing time. In the case of the Walker orbits with an inclination of $53^\circ$ the loss of observing time would be significantly lower and it would affect only the observations near the edges of the polar regions and in the SAA. A more precise estimation of the loss of observing time is currently ongoing.

\subsection{Radio communication and data transfer}
The satellite platform will use a UHF radio for house-keeping and telecommands, and we plan to integrate an S-band transmitter to downlink the scientific data. For GRB alerts, the payload will use a global inter-satellite communication network.

\begin{table}
\centering
\begin{tabular}{lrrl}
\hline
\hline
                           & Average Walker orbit & Average SSO orbit &            \\
\hline
\\[-2.4ex]
Total visibility time      & 42.40971             & 30.6859           & h          \\
Number of passes           & 212                  & 192               & passes     \\
Daily average pass number  & 6.8                  & 6.2               & passes/day \\
Average passes duration    & 12.00275             & 9.589343          & min        \\
Deviation of pass duration & 1.787037             & 2.180213          & min        \\
Maximum pass duration      & 13.5288              & 12.2411           & min        \\
Minimum pass duration      & 1.982317             & 0.769633          & min        \\
One visible satellite      & 10.10471             & 10.31914          & day        \\
Two visible satellites     & 0.777465             & 0.563032          & day        \\
Loss of visibility time    & 7.14                 & 5.17              & \%         \\
\hline
\\[-2.4ex]
\end{tabular}
\caption{Visibility time statistics for a 9-satellite constellation on two types of orbit. The assumed location of the ground station is in Budapest, Hungary. The analysis was performed for one typical month in the mission.}
\label{tab:visibility_time}
\end{table}

\begin{figure}[t]
\begin{center}
\includegraphics[width=0.99\linewidth]{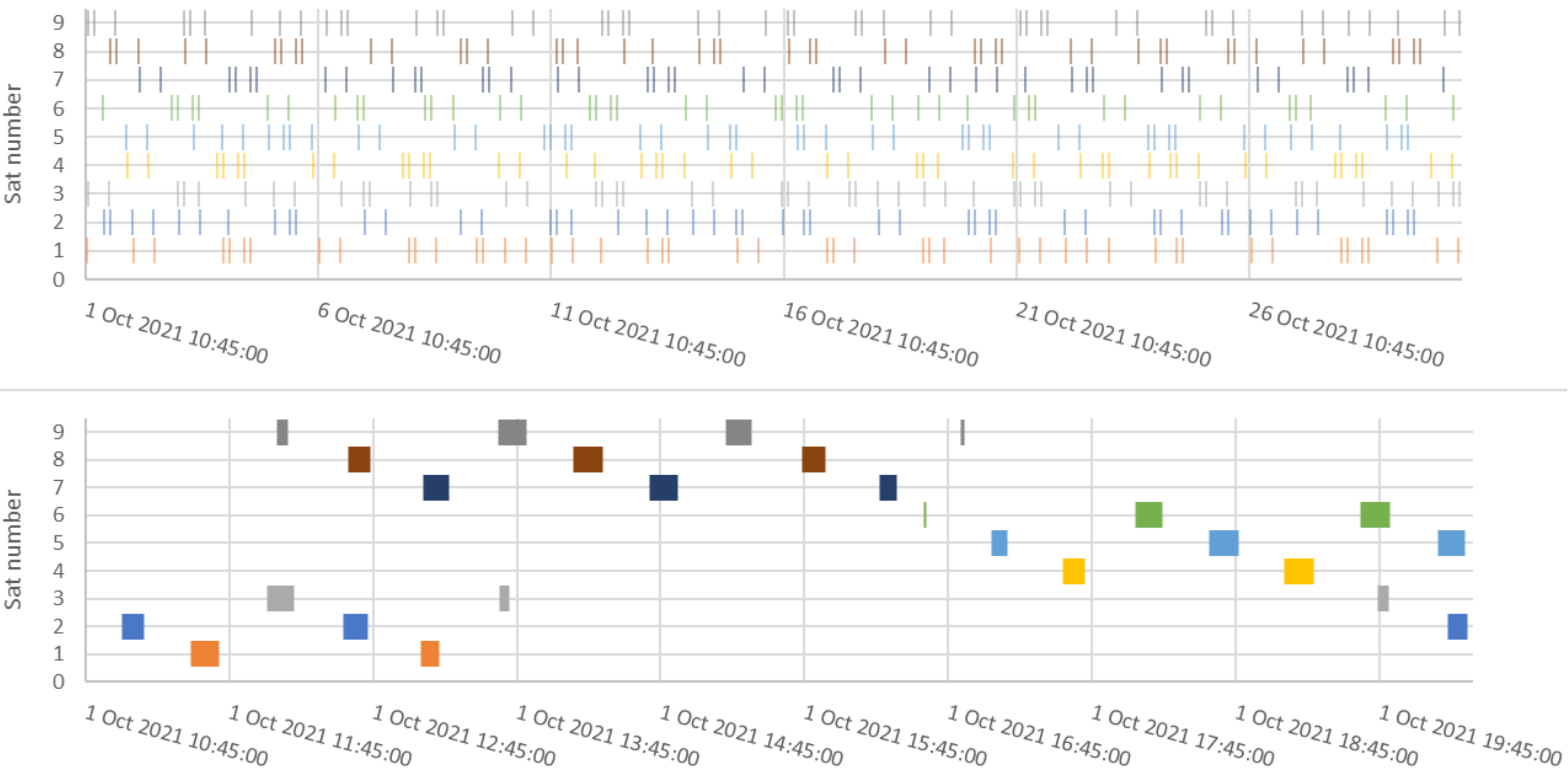}
\end{center}
\caption[example]
{ \label{fig:visibility_periods_polar}
Visibility periods for a constellation on SSO from Budapest for one month ({\it top}), for 10 hours ({\it bottom}).}
\end{figure}

We estimate the downlink data rates for a constellation of 9 satellites assuming Walker orbits and polar sun-synchronous orbits using one ground station in Budapest, Hungary. We only consider those passes, which are at least two minutes long. Table~\ref{tab:visibility_time} shows the visibility times for the two sets of orbits. Figure~\ref{fig:visibility_periods_polar} shows the timeline for the SSO constellation visibility from Budapest. There are only few overlapping passes.
The average downlink rate per satellite via S-band radio is 352.0 Mbyte/day for a constellation on Walker orbits and 359.5 Mbyte/day on sun-synchronous polar orbits. 
The daily amount of downlinked data for the whole constellation is 3168.3 Mbyte assuming Walker orbits and 3235.6 Mbyte for sun-synchronous polar orbits. Having two additional ground stations, for example one in East Asia and the other one in North America, would effectively triple the amount of downloadable data.

Three networks are currently available for Inter-Satellite Link (ISL) radios: Iridium/Iridium NEXT, Orblink and Globalstar. The highest data rate will be provided by Iridium-NEXT, which is still being deployed and is scheduled to be completed by the end of 2018. Iridium NEXT will have a data rate of 1.5 Mb/s.

\section{The expected detection rates and localization accuracy}
\label{sec:detect_rates_local}

We estimate the expected statistics of astrophysical bursts detected by the proposed constellation based on the detection rates of the Gamma-ray Burst Monitor (GBM) \cite{mee09} on the {\em Fermi} satellite \cite{atw94}. GBM consists of 12 sodium iodide and 2 bismuth germanate scintillators designed to detect gamma rays in the energy range from $\sim 8$\,keV to $\sim 40$\,MeV over the full sky except the region occulted by the Earth ({\it Fermi} GBM covers about 75\% of the sky). The sensitivity of the sodium iodide detectors is similar to the expected sensitivity of our proposed detector system and therefore the lessons learned by GBM are particularly useful for {\it CAMELOT}.

\begin{figure}[t]
\centering
\includegraphics[width=0.6\textwidth]{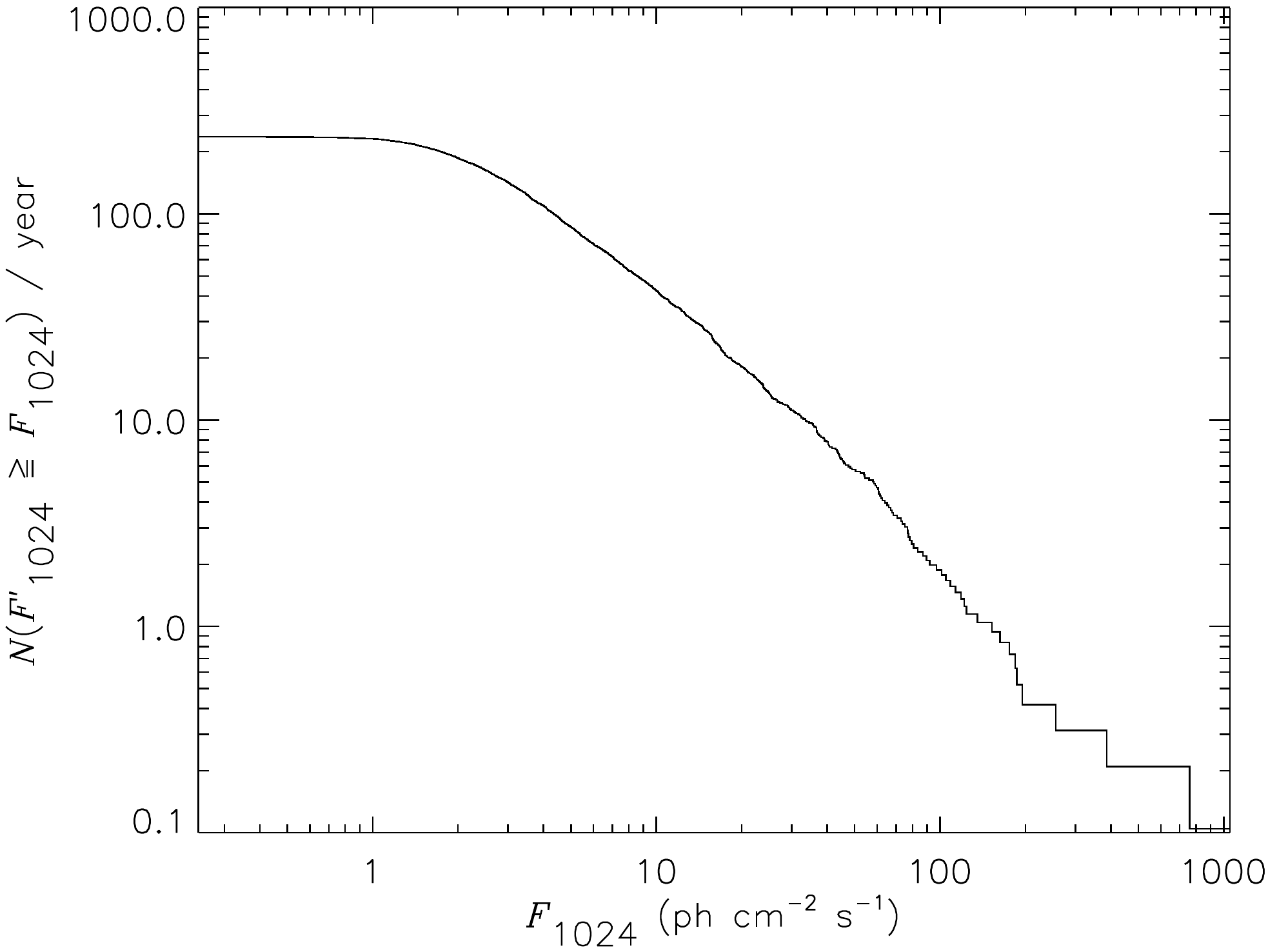}
\caption{The cumulative distribution function of the 1024-msec peak fluxes of all GRBs detected per year by {\em Fermi}/GBM. Each nanosatellite in the {\em CAMELOT} constellation will provide an effective area similar to two {\em Fermi}/GBM NaI detectors and therefore the constellation providing all-sky coverage is expected to detect about 1.3 times more GRBs than {\em Fermi}.}
\label{fig:gbm-peak_fluxes}
\end{figure}

Based on the Fermi GBM Burst Catalog\footnote{\url{https://heasarc.gsfc.nasa.gov/W3Browse/fermi/fermigbrst.html}} (FERMIGBRST), {\em Fermi}/GBM issued 2262 GRB triggers between 2008 July 12 and 2018 February 5 which is $\sim 0.65$ GRBs per day or $\sim 236$ per year. The number of SGRBs ($T_{90}$ duration $<2$\,s) is 39 per year and the number of LGRBs ($T_{90}$ duration $\geq 2$\,s) is 197 per year. The cumulative distribution function of the 1024-msec peak fluxes of all GRBs detected per year is shown in Figure~\ref{fig:gbm-peak_fluxes}. Table~\ref{tab:gbm-peak_fluxes} gives the numbers of GRBs detected per year in several flux bins.

The effective area of the two-sided perpendicular detector configuration considered for {\em CAMELOT} (maximum $\sim 340$\,cm$^2$) is similar to two {\em Fermi}/GBM NaI detectors. Therefore the expected trigger rate of GRBs for one cubesat will be similar to {\em Fermi}/GBM, assuming the same orbit. If we use polar orbits, the observing efficiency will be reduced by $\sim 30$\,\% to $\sim 40$\,\% as shown in Section~\ref{background_estimation} and the trigger rate will also be reduced by a similar amount. For the all-sky coverage provided by the full constellation, the expected total number of detected GRBs is $\sim310$ per year, $\sim1.3$ times more than the number of GRBs detected by {\it Fermi} GBM.

\begin{table}[h]
\centering
\begin{tabular}{lrrrrrrr}
\hline
\hline
\\[-2.4ex]
$F_{1024}$ (ph\,cm$^{-2}$\,s$^{-1}$) &   1 &   3 &  5 & 10 & 30 & 50 & 100 \\
$N$/year                             & 230 & 142 & 86 & 42 & 11 &  6 &   2 \\
\hline
\\[-2.4ex]
\end{tabular}
\caption{Number of GRBs detected by {\em Fermi}/GBM per year ($N$/year) with 1024-msec peak flux ($F_{1024}$) higher or equal than a given value. The detection rate of the full {\it CAMELOT} constellation is expected to be about 1.3 times higher than that of {\em Fermi}/GBM.}
\label{tab:gbm-peak_fluxes}
\end{table}

The {\it Fermi} GBM Trigger Catalog\footnote{\url{https://heasarc.gsfc.nasa.gov/W3Browse/fermi/fermigtrig.html}} (FERMIGTRIG), shows that {\em Fermi}/GBM also often triggers to other transient sources. The frequency of various kinds of triggers between 2008 July 12 and 2018 February 5 is summarized in Table~\ref{tab:gbm-triggers}. The total number of all triggers in the catalog, including GRBs, is 6211. Non-GRB sources make up 63\,\% of all issued triggers. The trigger types in the catalog are: SF - Solar flare; LOCPAR - Local particles; TGF - Terrestrial gamma-ray flashes (extremely short, sub-millisecond to milliseconds, spectrally hard Earth's atmospheric bursts associated with lightning events in thunderstorms); TRANS - Generic transient; UNCERT - Uncertain classification; SGR - Soft gamma repeater; DISTPAR - Distance particle event; GALBIN - Galactic binary. These observations will provide important data for exciting secondary science. 

\begin{table}[h]
\centering
\begin{tabular}{ccccccccc}
\hline
\hline
\\[-2.4ex]
Type       & SF   & LOCPAR & TGF & TRANS & UNCERT & SGR & DISTPAR & GALBIN \\
\hline
\\[-2.4ex]
$N$        & 1175 & 924    & 830 & 367  & 316     & 253 & 76      & 3      \\
$N$/year   & 123  & 97     & 87  & 38   & 33      & 26  & 8       & 0.3    \\
Frac. (\%) & 19   & 15     & 13  & 6    & 5       & 4   & 1       & $<0.1$ \\
\hline
\\[-2.4ex]
\end{tabular}
\caption{Summary of the number $N$ of various kinds of triggers, other than GRBs, between 2008 July 12 and 2018 February 5 in the Fermi GBM Trigger Catalog. Frac. is the fraction of the given type of trigger within all 6211 triggers in the catalog. Non-GRB sources make up 63\,\% of all issued triggers. Similarly to {\it Fermi} GBM, {\it CAMELOT} will also detect many non-GRB transients which will provide opportunity for exciting secondary science.}
\label{tab:gbm-triggers}
\end{table}

\begin{figure}[t]
\begin{center}
\includegraphics[width=0.8\linewidth]{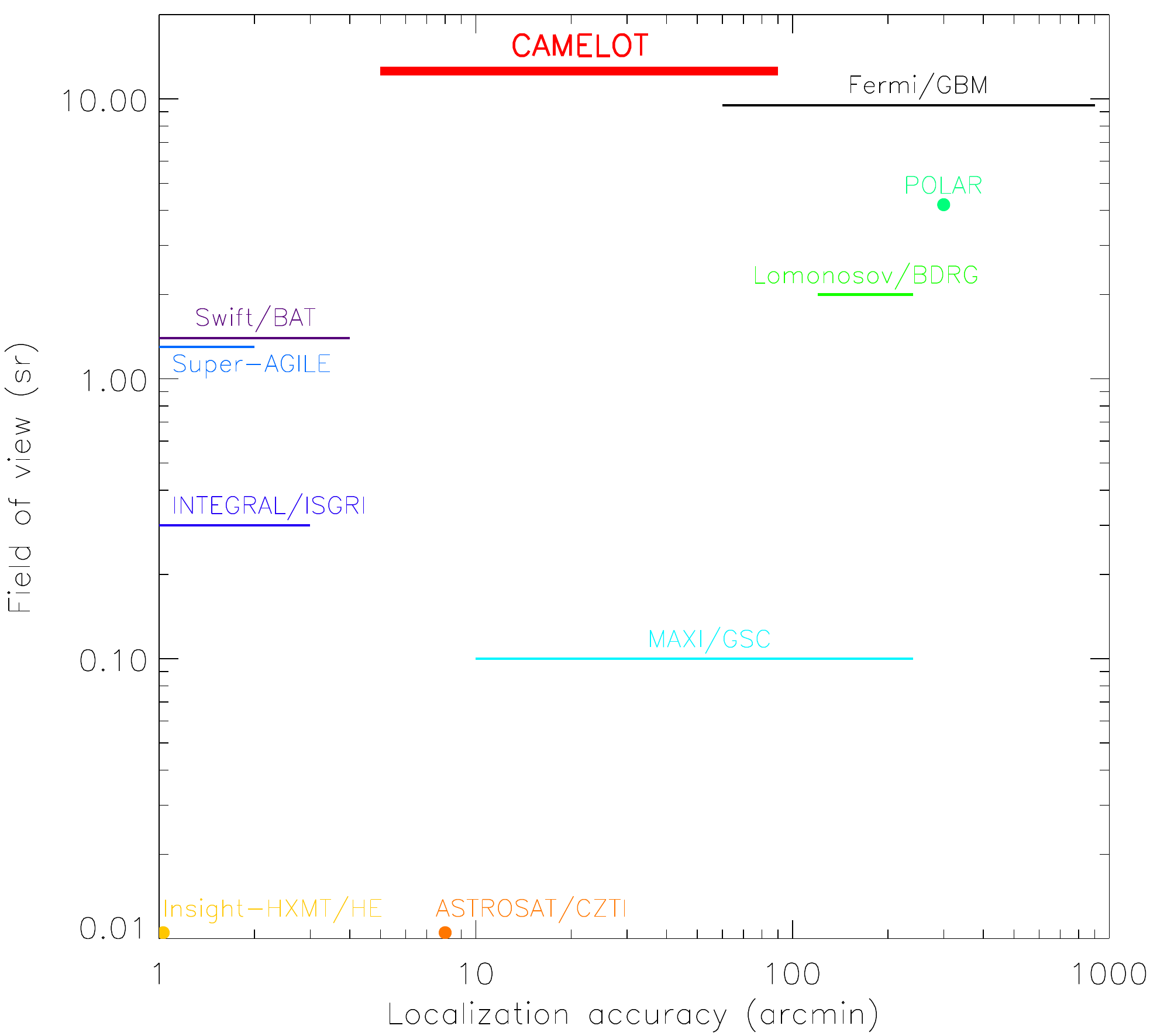}
\end{center}
\caption[example]
{ \label{fig:loc-acc-fov-overview}
Field of view vs. localization accuracy for the currently operating GRB monitoring instruments. By providing both all-sky coverage and good localization accuracy, the proposed {\it CAMELOT} mission fills an empty region in the parameter space.}
\end{figure}

According to The Third {\em Fermi} GBM Gamma-Ray Burst Catalog: The First Six Years \cite{nar16}, the GBM instrument triggered during the period from 2008 July 12 to 2014 July 11 approximately in 10\,\% of cases due to cosmic rays or trapped particles; the latter typically occur in the entry or exit regions of the SAA or at high geomagnetic latitudes. In rare cases, outbursts from known Galactic sources also caused triggers. About 6\,\% of the GBM triggers are generated accidentally by statistical fluctuations or are too weak to be confidently classified. 

Our project will benefit from the all-sky coverage and high localization accuracy (see Figure~\ref{fig:loc-acc-fov-overview}). The expected localization accuracy will range from $\sim 10^\prime$ to a few degrees depending on the flux and duration of a GRB (see Ohno et al. this conference).  Current missions have either large FoV or good localization accuracy, but {\it none of the current GRB missions provides both all-sky coverage and source localization}. The large FoV instruments, e.g. {\em INTEGRAL}/SPI-ACS \cite{ved03} or {\em Fermi}/GBM \cite{mee09}, have either no localization capability or the localization accuracy is of the order of several degrees. On the other hand the current instruments providing localization accuracy of the order of arcmin have limited FoV, e.g. {\em Swift}/BAT \cite{bar05} or {\em AGILE}/Super-AGILE \cite{fer07}.

{\em CAMELOT} is expected to provide data and rapid alerts to the currently operating GRB collection and alert networks such as: I. the Gamma-ray Coordinates Network (Transient Astronomy Network) (GCN/TAN)\footnote{\url{https://gcn.gsfc.nasa.gov}} which provides information about GRBs (locations, images, spectra, lightcurves, follow-up observation reports) in real-time to the world community; II. the Inter Planetary Network (IPN)\footnote{\url{http://www.ssl.berkeley.edu/ipn3}} which derives the positions of fast gamma-ray transients of all kinds by triangulation; III. the Global Relay of Observatories Watching Transients Happen (GROWTH)\footnote{\url{http://growth.caltech.edu/index.html}} which is an international scientific collaborative project studying astronomical transients. These alerts will allow quick follow-up observations by many existing and future ground based observatories, such as the Mobile Astronomical System of TElescope Robots (MASTER); the Burst Optical Observer and Transient Exploring System (BOOTES); the Robotic Optical Transient Search Experiment (ROTSE); Pi of the Sky, which perform photometric or spectroscopic observations of GRB afterglows.

Several other nanosatellite missions have been proposed for GRB observations, such as NASA's {\it burstCube} \cite{racusin2017}, the Italian {\it Hermes} constellation (Fiore et al. this conference), the Chinese {\it GRID}, and others. These various missions provide ample opportunity for collaboration. By working together, sharing data and ground stations, the various nanosatellite missions can extend each other's capabilities and further improve the rapid localization of gamma-ray transients.

\section{Summary}

The proposed constellation of nanosatellites  will provide all-sky coverage with high sensitivity and localization accuracy, as well as rapid data downlink following detections of $\gamma$-ray transients. 

We propose a constellation of 3U cubesats equipped with CsI scintillator based gamma-ray detectors, read out by multi-pixel photon counters (see Ohno et al. this conference). Each nanosatellite shall carry four thin, 9\,mm, and relatively large, $8.3\times15$\,cm, detectors as lateral extensions on its surface, on two perpendicular sides. The large thin detectors provide high sensitivity, while leaving enough room for electronics. 

In terms of all-sky monitoring, the proposed fleet will outperform all past and existing GRB monitoring missions. Nine satellites in low-Earth Walker orbits (500-600 km), in three orbital planes with inclinations of $53^\circ$, will simultaneously cover 95\% of the sky by five satellites. By cross-correlating the light curves of the detected GRBs, the fleet shall be able to determine the time difference of the arriving signal between the satellites and thus determine the position of bright short bursts with an accuracy $\sim 10^\prime$. This requirement demands precise time synchronization between the satellites and accurate time stamping of the detected gamma-ray photons. This will be achieved by using miniaturized space borne GPS receivers (see P\'al et al. this conference). 

Additionally, the approximate cosine dependence of the effective area with incidence angle will allow to localize GRBs by measuring the relative brightness of the burst detected by different detectors to an accuracy potentially as good as a few degrees. This will be particularly important for the localization of faint GRBs.

The sources of short GRBs are among the most important sources of gravitational waves detected with LIGO. LIGO has a modest localization accuracy, limiting our knowledge about these astrophysically extremely important events. Simultaneous detections of GWs and GRBs, with accurately measured locations in the sky, will therefore be important for enabling follow up observations, providing valuable multi-messenger information about these exciting events.

Rapid follow up observations at other wavelengths require the capability for fast, nearly simultaneous downlink of data for the triggered events from all satellites in the fleet. This can be achieved using satellite-to-satellite communication networks such as Iridium NEXT.

The same payload will also provide important secondary science, such as monitoring of outbursts of soft gamma-ray repeaters, gamma-ray flares on the Sun, terrestrial gamma-ray flashes (produced in thunderstorms), and space weather phenomena, such as monitoring of particles in low Earth orbit.

This constellation of satellites is a mission which provides ample potential for international cooperation. Because the proposed fleet is scalable and extendable, we also envision future partners joining with different satellite designs, potentially extending the capabilities of the originally proposed constellation.

\acknowledgments 
This work was supported by the Lend\"uet LP2016-11 and LP2012-31 grants awarded by the Hungarian Academy of Sciences, by GINOP-2.3.2-15-2016-00033, as well as by Hiroshima university, JSPS KAKENHI Grant Number
17H06362.

\bibliography{report} 

\begin{thebibliography}{10}

\bibitem{pir04}
{Piran}, T., ``{The physics of gamma-ray bursts},'' {\em Reviews of Modern
  Physics}~{\bf 76},  1143--1210 (2004).

\bibitem{mesp06}
{M{\'e}sz{\'a}ros}, P., ``{Gamma-ray bursts},'' {\em Reports on Progress in
  Physics}~{\bf 69},  2259--2321 (2006).

\bibitem{ved09}
{Vedrenne}, G. and {Atteia}, J.-L.,  [{\em {Gamma-Ray Bursts: The brightest
  explosions in the Universe}}{\nolinebreak\hspace{0.1em}]}, Springer and
  Praxis Publishing Ltd, Chichester (2009).

\bibitem{kou12}
{Kouveliotou}, C., {Wijers}, R.~A.~M.~J., and {Woosley}, S.,  [{\em {Gamma-ray
  Bursts}}{\nolinebreak\hspace{0.1em}]}, Cambridge University Press, Cambridge
  (2012).

\bibitem{ger12}
{Gehrels}, N. and {M{\'e}sz{\'a}ros}, P., ``{Gamma-Ray Bursts},'' {\em
  Science}~{\bf 337},  932 (2012).

\bibitem{gom12}
{Gomboc}, A., ``{Unveiling the secrets of gamma ray bursts},'' {\em
  Contemporary Physics}~{\bf 53},  339--355 (2012).

\bibitem{ger13}
{Gehrels}, N. and {Razzaque}, S., ``{Gamma-ray bursts in the swift-Fermi
  era},'' {\em Frontiers of Physics}~{\bf 8},  661--678 (2013).

\bibitem{kum15}
{Kumar}, P. and {Zhang}, B., ``{The physics of gamma-ray bursts and
  relativistic jets},'' {\em Physics Reports}~{\bf 561},  1--109 (2015).

\bibitem{wil17}
{Willingale}, R. and {M{\'e}sz{\'a}ros}, P., ``{Gamma-Ray Bursts and Fast
  Transients. Multi-wavelength Observations and Multi-messenger Signals},''
  {\em Space Science Reviews}~{\bf 207},  63--86 (2017).

\bibitem{kle73}
{Klebesadel}, R.~W., {Strong}, I.~B., and {Olson}, R.~A., ``{Observations of
  Gamma-Ray Bursts of Cosmic Origin},'' {\em The Astrophysical Journal
  Letters}~{\bf 182},  L85 (1973).

\bibitem{maz74}
{Mazets}, E.~P., {Golenetskij}, S.~V., and {Il'Inskij}, V.~N., ``{Burst of
  cosmic gamma -emission from observations on Cosmos 461.},'' {\em Pisma v
  Zhurnal Eksperimentalnoi i Teoreticheskoi Fiziki}~{\bf 19},  126--128 (1974).

\bibitem{kou93}
{Kouveliotou}, C., {Meegan}, C.~A., {Fishman}, G.~J., {Bhat}, N.~P., {Briggs},
  M.~S., {Koshut}, T.~M., {Paciesas}, W.~S., and {Pendleton}, G.~N.,
  ``{Identification of two classes of gamma-ray bursts},'' {\em The
  Astrophysical Journal Letters}~{\bf 413},  L101--L104 (1993).

\bibitem{maz81b}
{Mazets}, E.~P., {Golenetskii}, S.~V., {Ilinskii}, V.~N., {Panov}, V.~N.,
  {Aptekar}, R.~L., {Gurian}, I.~A., {Proskura}, M.~P., {Sokolov}, I.~A.,
  {Sokolova}, Z.~I., and {Kharitonova}, T.~V., ``{Catalog of cosmic gamma-ray
  bursts from the KONUS experiment data. I.},'' {\em Astrophysics and Space
  Science}~{\bf 80},  3--83 (1981).

\bibitem{bal03}
{Bal{\'a}zs}, L.~G., {Bagoly}, Z., {Horv{\'a}th}, I., {M{\'e}sz{\'a}ros}, A.,
  and {M{\'e}sz{\'a}ros}, P., ``{On the difference between the short and long
  gamma-ray bursts},'' {\em Astronomy and Astrophysics}~{\bf 401},  129--140
  (2003).

\bibitem{bor04}
{Borgonovo}, L., ``{Bimodal distribution of the autocorrelation function in
  gamma-ray bursts},'' {\em Astronomy and Astrophysics}~{\bf 418},  487--493
  (2004).

\bibitem{mesa06}
{M{\'e}sz{\'a}ros}, A., {Bagoly}, Z., {Bal{\'a}zs}, L.~G., and {Horv{\'a}th},
  I., ``{Redshift distribution of gamma-ray bursts and star formation rate},''
  {\em Astronomy and Astrophysics}~{\bf 455},  785--790 (2006).

\bibitem{zha09}
{Zhang}, B., {Zhang}, B.-B., {Virgili}, F.~J., {Liang}, E.-W., {Kann}, D.~A.,
  {Wu}, X.-F., {Proga}, D., {Lv}, H.-J., {Toma}, K., {M{\'e}sz{\'a}ros}, P.,
  and et~al., ``{Discerning the Physical Origins of Cosmological Gamma-ray
  Bursts Based on Multiple Observational Criteria: The Cases of z = 6.7 GRB
  080913, z = 8.2 GRB 090423, and Some Short/Hard GRBs},'' {\em The
  Astrophysical Journal}~{\bf 703},  1696--1724 (2009).

\bibitem{ber14}
{Berger}, E., ``{Short-Duration Gamma-Ray Bursts},'' {\em Annual Review of
  Astronomy and Astrophysics}~{\bf 52},  43--105 (2014).

\bibitem{abb17a}
{Abbott}, B.~P., {Abbott}, R., {Abbott}, T.~D., {Acernese}, F., {Ackley}, K.,
  {Adams}, C., {Adams}, T., {Addesso}, P., {Adhikari}, R.~X., {Adya}, V.~B.,
  and et~al., ``{Gravitational Waves and Gamma-Rays from a Binary Neutron Star
  Merger: GW170817 and GRB 170817A},'' {\em The Astrophysical Journal
  Letters}~{\bf 848},  L13 (2017).

\bibitem{abb17b}
{Abbott}, B.~P., {Abbott}, R., {Abbott}, T.~D., {Acernese}, F., {Ackley}, K.,
  {Adams}, C., {Adams}, T., {Addesso}, P., {Adhikari}, R.~X., {Adya}, V.~B.,
  and et~al., ``{GW170817: Observation of Gravitational Waves from a Binary
  Neutron Star Inspiral},'' {\em Physical Review Letters}~{\bf 119}(16),
  161101 (2017).

\bibitem{abb17c}
{Abbott}, B.~P., {Abbott}, R., {Abbott}, T.~D., {Acernese}, F., {Ackley}, K.,
  {Adams}, C., {Adams}, T., {Addesso}, P., {Adhikari}, R.~X., {Adya}, V.~B.,
  and et~al., ``{Multi-messenger Observations of a Binary Neutron Star
  Merger},'' {\em The Astrophysical Journal Letters}~{\bf 848},  L12 (2017).

\bibitem{tan13}
{Tanvir}, N.~R., {Levan}, A.~J., {Fruchter}, A.~S., {Hjorth}, J., {Hounsell},
  R.~A., {Wiersema}, K., and {Tunnicliffe}, R.~L., ``{A `kilonova' associated
  with the short-duration {$\gamma$}-ray burst GRB 130603B},'' {\em
  Nature}~{\bf 500},  547--549 (2013).

\bibitem{kag17}
{KAGRA Collaboration}, {Akutsu}, T., {Ando}, M., {Araya}, A., {Aritomi}, N.,
  {Asada}, H., {Aso}, Y., {Atsuta}, S., {Awai}, K., {Barton}, M.~A., {Cannon},
  K., {Craig}, K., {Creus}, W., {Doi}, K., and et~al., ``{The status of KAGRA
  underground cryogenic gravitational wave telescope},'' {\em arXiv:1710.04823}
   (2017).

\bibitem{vet91}
{Vette}, J.~I., ``{The AE-8 trapped electron model environment},'' {\em NASA
  STI Technical Report, NSSDC/WDC-A-R\&S 91-24}  (1991).

\bibitem{vam96}
{Vampola}, A.~L., ``{The ESA Outer Zone Electron Model Update},'' in [{\em
  Environment Modeling for Space-Based
  Applications}{\nolinebreak\hspace{0.1em}]},  {Guyenne}, T.-D. and {Hilgers},
  A., eds., {\em ESA Special Publication} {\bf 392},  151 (1996).

\bibitem{saw76}
{Sawyer}, D.~M. and {Vette}, J.~I., ``{AP-8 trapped proton environment for
  solar maximum and solar minimum},'' {\em NASA STI Technical Report,
  NSSDC/WDC-A-R/S-76-06}  (1976).

\bibitem{sad17}
{Sadovnichii}, V.~A., {Panasyuk}, M.~I., {Amelyushkin}, A.~M., {Bogomolov},
  V.~V., {Benghin}, V.~V., {Garipov}, G.~K., {Kalegaev}, V.~V., {Klimov},
  P.~A., {Khrenov}, B.~A., {Petrov}, V.~L., and et~al., ``{``Lomonosov''
  Satellite--Space Observatory to Study Extreme Phenomena in Space},'' {\em
  Space Science Reviews}~{\bf 212},  1705--1738 (2017).

\bibitem{sve18}
{Svertilov}, S.~I., {Panasyuk}, M.~I., {Bogomolov}, V.~V., {Amelushkin}, A.~M.,
  {Barinova}, V.~O., {Galkin}, V.~I., {Iyudin}, A.~F., {Kuznetsova}, E.~A.,
  {Prokhorov}, A.~V., {Petrov}, V.~L., and et~al., ``{Wide-Field
  Gamma-Spectrometer BDRG: GRB Monitor On-Board the Lomonosov Mission},'' {\em
  Space Science Reviews}~{\bf 214},  \#8 (2018).

\bibitem{ben18}
{Benghin}, V.~V., {Nechaev}, O.~Y., {Zolotarev}, I.~A., {Amelyushkin}, A.~M.,
  {Petrov}, V.~L., {Panasyuk}, M.~I., and {Yashin}, I.~V., ``{An Experiment in
  Radiation Measurement Using the Depron Instrument},'' {\em Space Science
  Reviews}~{\bf 214},  \#9 (2018).

\bibitem{mee09}
{Meegan}, C., {Lichti}, G., {Bhat}, P.~N., {Bissaldi}, E., {Briggs}, M.~S.,
  {Connaughton}, V., {Diehl}, R., {Fishman}, G., {Greiner}, J., {Hoover},
  A.~S., and et~al., ``{The Fermi Gamma-ray Burst Monitor},'' {\em The
  Astrophysical Journal}~{\bf 702},  791--804 (2009).

\bibitem{atw94}
{Atwood}, W.~B. and {GLAST Collaboration}, ``{Gamma Large Area Silicon
  Telescope (GLAST) applying silicon strip detector technology to the detection
  of gamma rays in space},'' {\em Nuclear Instruments and Methods in Physics
  Research A}~{\bf 342},  302--307 (1994).

\bibitem{nar16}
{Narayana Bhat}, P., {Meegan}, C.~A., {von Kienlin}, A., {Paciesas}, W.~S.,
  {Briggs}, M.~S., {Burgess}, J.~M., {Burns}, E., {Chaplin}, V., {Cleveland},
  W.~H., {Collazzi}, A.~C., and et~al., ``{The Third Fermi GBM Gamma-Ray Burst
  Catalog: The First Six Years},'' {\em The Astrophysical Journal Supplement
  Series}~{\bf 223},  28 (2016).

\bibitem{ved03}
{Vedrenne}, G., {Roques}, J.-P., {Sch{\"o}nfelder}, V., {Mandrou}, P.,
  {Lichti}, G.~G., {von Kienlin}, A., {Cordier}, B., {Schanne}, S.,
  {Kn{\"o}dlseder}, J., {Skinner}, G., and et~al., ``{SPI: The spectrometer
  aboard INTEGRAL},'' {\em Astronomy and Astrophysics}~{\bf 411},  L63--L70
  (2003).

\bibitem{bar05}
{Barthelmy}, S.~D., {Barbier}, L.~M., {Cummings}, J.~R., {Fenimore}, E.~E.,
  {Gehrels}, N., {Hullinger}, D., {Krimm}, H.~A., {Markwardt}, C.~B., {Palmer},
  D.~M., {Parsons}, A., and et~al., ``{The Burst Alert Telescope (BAT) on the
  SWIFT Midex Mission},'' {\em Space Science Reviews}~{\bf 120},  143--164
  (2005).

\bibitem{fer07}
{Feroci}, M., {Costa}, E., {Soffitta}, P., {Del Monte}, E., {di Persio}, G.,
  {Donnarumma}, I., {Evangelista}, Y., {Frutti}, M., {Lapshov}, I.,
  {Lazzarotto}, F., {Mastropietro}, M., {Morelli}, E., {Pacciani}, L.,
  {Porrovecchio}, G., and et~al., ``{SuperAGILE: The hard X-ray imager for the
  AGILE space mission},'' {\em Nuclear Instruments and Methods in Physics
  Research A}~{\bf 581},  728--754 (2007).

\bibitem{racusin2017}
{Racusin}, J., {Perkins}, J.~S., {Briggs}, M.~S., {de Nolfo}, G., {Krizmanic},
  J., {Caputo}, R., {McEnery}, J.~E., {Shawhan}, P., {Morris}, D.,
  {Connaughton}, V., {Kocevski}, D., {Wilson-Hodge}, C., {Hui}, M., {Mitchell},
  L., and {McBreen}, S., ``{BurstCube: A CubeSat for Gravitational Wave
  Counterparts},'' {\em ArXiv e-prints}  (Aug. 2017).

\end{thebibliography}
\bibliographystyle{spiebib} 

\end{document}